\documentclass[10pt,journal,compsoc]{IEEEtran}

\ifCLASSINFOpdf
  \pdfoutput=1\relax                   
  \usepackage{graphicx}                
  \DeclareGraphicsExtensions{.pdf,.png,.jpg,.jpeg,.eps} 
\else
  \usepackage{graphicx}                
  \DeclareGraphicsExtensions{.eps}     
\fi%

\graphicspath{{figures/}{pictures/}{images/}{./}} 

\usepackage{amsmath}
\usepackage{amssymb}
\usepackage{mathtools}
\usepackage{bbm}
\usepackage{microtype}                 
\PassOptionsToPackage{warn}{textcomp}  
\usepackage{textcomp}                  
\usepackage{mathptmx}                  
\usepackage{times}                     

\usepackage{array}
\usepackage{scalerel}
\usepackage{tikz}
\usepackage{bm}
\usepackage{xcolor}
\usepackage{colortbl}
\usepackage[percent]{overpic}
\usepackage{wrapfig}
\usepackage{multirow}
\usepackage[export]{adjustbox}
\usepackage{caption}
\ifCLASSOPTIONcompsoc
  \usepackage[caption=false,font=footnotesize,labelfont=sf,textfont=sf]{subfig}
\else
  \usepackage[caption=false,font=footnotesize]{subfig}
\fi
\usepackage{stfloats}
\usepackage{tabularx}
\usepackage[figuresleft]{rotating}
\usepackage{hyperref}
\usepackage{url}

\usepackage{algpseudocode}
\usepackage{algorithm}
\usepackage{comment}

\usetikzlibrary{shapes,arrows,decorations.pathreplacing}
\usetikzlibrary{svg.path}
\definecolor{orcidlogocol}{HTML}{A6CE39}
\tikzset{
  orcidlogo/.pic={
    \fill[orcidlogocol] svg{M256,128c0,70.7-57.3,128-128,128C57.3,256,0,198.7,0,128C0,57.3,57.3,0,128,0C198.7,0,256,57.3,256,128z};
    \fill[white] svg{M86.3,186.2H70.9V79.1h15.4v48.4V186.2z}
                 svg{M108.9,79.1h41.6c39.6,0,57,28.3,57,53.6c0,27.5-21.5,53.6-56.8,53.6h-41.8V79.1z M124.3,172.4h24.5c34.9,0,42.9-26.5,42.9-39.7c0-21.5-13.7-39.7-43.7-39.7h-23.7V172.4z}
                 svg{M88.7,56.8c0,5.5-4.5,10.1-10.1,10.1c-5.6,0-10.1-4.6-10.1-10.1c0-5.6,4.5-10.1,10.1-10.1C84.2,46.7,88.7,51.3,88.7,56.8z};
  }
}
\newcommand\orcidicon[1]{\href{https://orcid.org/#1}{\mbox{\scalerel*{
\begin{tikzpicture}[yscale=-1,transform shape]
\pic{orcidlogo};
\end{tikzpicture}
}{|}}}}

\usepackage{filecontents}

\usepackage[numbers]{natbib}
\setcitestyle{open={[},close={]},citesep={,}}
\usepackage[resetlabels]{multibib}
\newcites{body}{References}
\newcites{app}{References}

\newcommand{\etAl}{\textit{et~al.}}
\newcommand{\pluseq}{\mathrel{+}=}

\newlength{\mycaptionspacinglength}
\newcommand{\mycaptionspacing}{\vspace*{\mycaptionspacinglength}}
\newcommand{\R}{\mathbb{R}}

\newlength{\Oldarrayrulewidth}
\newcommand{\Cline}[2]{%
  \noalign{\global\setlength{\Oldarrayrulewidth}{\arrayrulewidth}}%
  \noalign{\global\setlength{\arrayrulewidth}{#1}}\cline{#2}%
  \noalign{\global\setlength{\arrayrulewidth}{\Oldarrayrulewidth}}}

\definecolor{changedcolor}{rgb}{0.188, 0.192, 0.874}

\newcommand{\changed}[1]{#1}

\newif\ifteaser
\teasertrue 

\begin{document}

\title{Learning Adaptive Sampling and Reconstruction\\for Volume Visualization}



\author{Sebastian~Weiss\orcidicon{0000-0003-4399-3180}, Mustafa~I\c{s}{\i}k\orcidicon{https://orcid.org/0000-0002-3086-8922}, Justus~Thies\orcidicon{https://orcid.org/0000-0002-0056-9825}, and R\"udiger~Westermann\orcidicon{https://orcid.org/0000-0002-3394-0731}%
\IEEEcompsocitemizethanks{\IEEEcompsocthanksitem%
All authors are with Technical University of Munich, Germany.\protect\\E-mail: \{sebastian13.weiss,m.isik, justus.thies, westermann\}@tum.de.}}


\IEEEtitleabstractindextext{%
\begin{abstract}
A central challenge in data visualization is to understand which data samples are required to generate an image of a data set in which the relevant information is encoded. In this work, we make a first step towards answering the question of whether an artificial neural network can predict where to sample the data with higher or lower density, by learning of correspondences between the data, the sampling patterns and the generated images. We introduce a novel neural rendering pipeline, which is trained end-to-end to generate a sparse adaptive sampling structure from a given low-resolution input image, and reconstructs a high-resolution image from the sparse set of samples. For the first time, to the best of our knowledge, we demonstrate that the selection of structures that are relevant for the final visual representation can be jointly learned together with the reconstruction of this representation from these structures. Therefore, we introduce differentiable sampling and reconstruction stages, which can leverage back-propagation based on supervised losses solely on the final image. We shed light on the adaptive sampling patterns generated by the network pipeline and analyze its use for volume visualization including isosurface and direct volume rendering.
\end{abstract}
}

\ifteaser
\newcommand{\showTeaser}{%
\begin{center}%
  \vspace{-20pt}
  \includegraphics[width=0.9\linewidth]{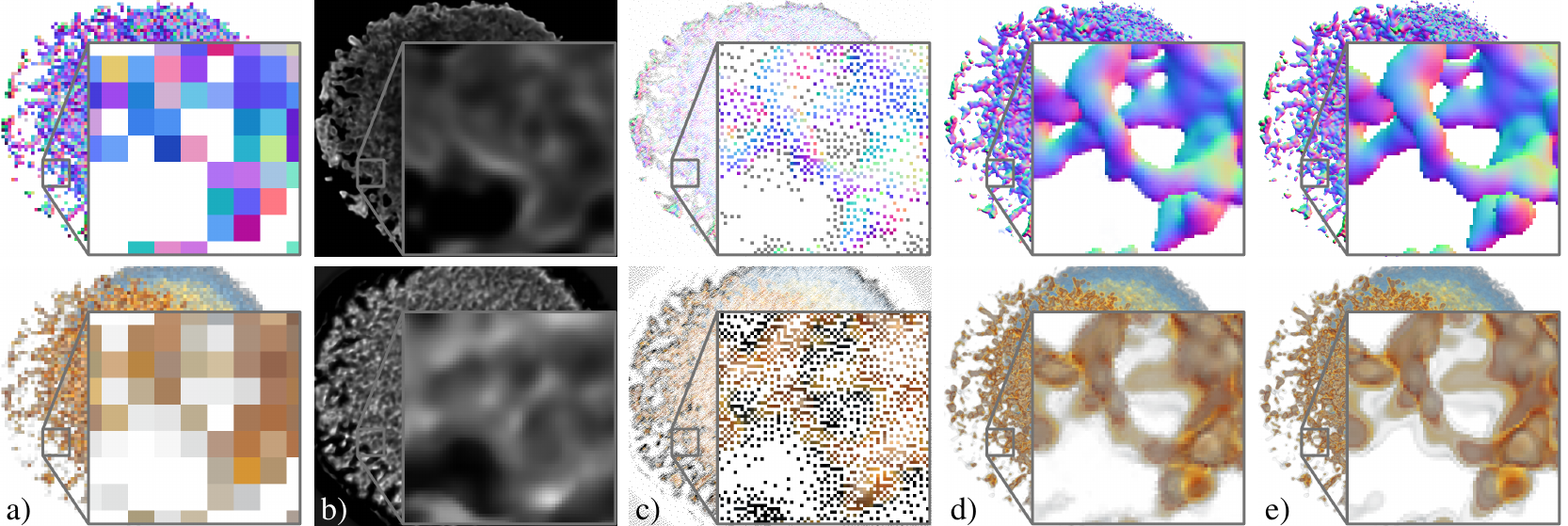}
  \setcounter{figure}{0}
  \captionof{figure}{An importance network, together with a differentiable sampler and a reconstruction network, takes a low resolution visualization (a) and infers an importance map (b) from it. From this map, an adaptive sampling pattern with adjustable number of samples ($5\%$ \changed{for iso, top; $10\%$ for dvr, bottom}) is derived, and a volume ray-caster samples the data according to these samples (c). The reconstruction network completes the visual representation from the sparse set of samples (d). The ground truth visualizations are shown in (e). The proposed network pipeline works on images of iso-surfaces (top) and direct volume renderings (bottom).}
  \label{fig:teaser}
  \vspace{20pt}
\end{center}
}
\else
\newcommand{\showTeaser}{}
\fi

\maketitle

\IEEEdisplaynontitleabstractindextext

%
\IEEEpeerreviewmaketitle

\ifteaser\else
\begin{figure*}
  \centering
  \includegraphics[width=0.9\linewidth]{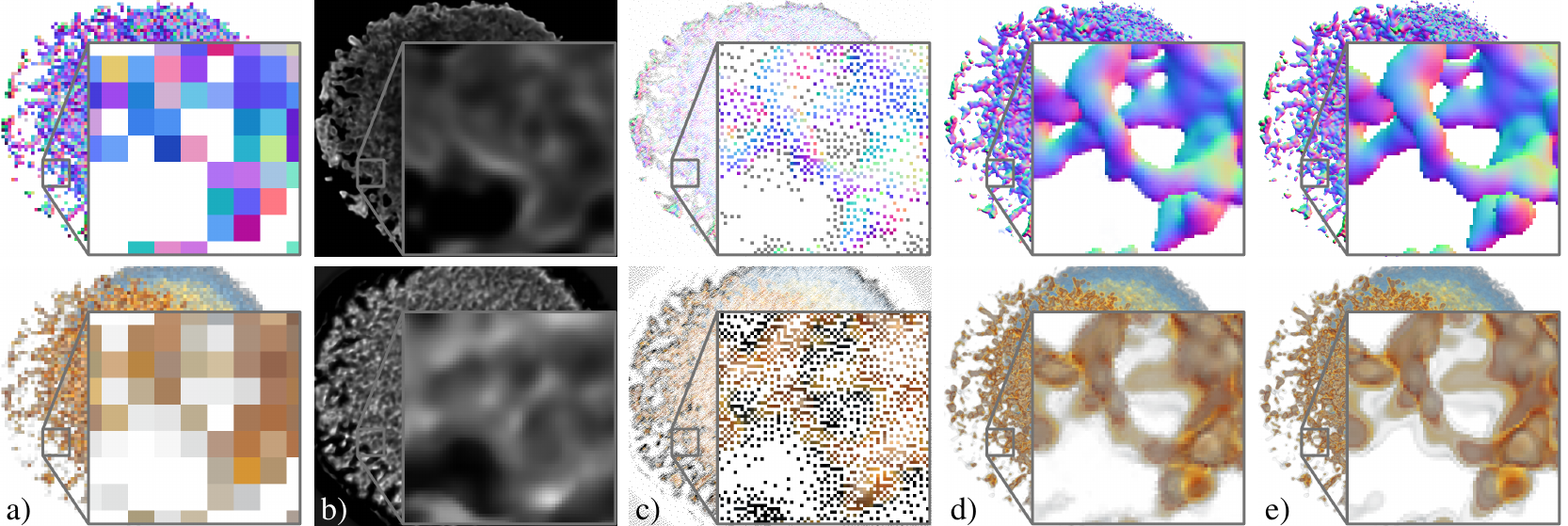}
  \caption{An importance network, together with a differentiable sampler and a reconstruction network, takes a low resolution visualization (a) and infers an importance map (b) from it. From this map, an adaptive sampling pattern with adjustable number of samples ($5\%$ \changed{for iso, top; $10\%$ for dvr, bottom}) is derived, and a volume ray-caster samples the data according to these samples (c). The reconstruction network completes the visual representation from the sparse set of samples (d). The ground truth visualizations are shown in (e). The proposed network pipeline works on images of iso-surfaces (top) and direct volume renderings (bottom).}
  \label{fig:teaser}
\end{figure*}
\fi


\IEEEraisesectionheading{\section{Introduction}\label{sec:introduction}}


\IEEEPARstart{W}{hich}
are the data samples that are needed to generate an image of a data set that conveys the relevant information encoded in this data? This question is fundamental to data visualization, since it asks for the importance of data samples from a perceptual point of view, rather than a signal processing standpoint that argues in terms of numerical accuracy.  

Recent works in visualization have shown that artificial neural networks can perform an accurate reconstruction from a reduced set of data samples, by learning the relationships between a sparse, yet regular input sampling and the high-resolution output. Learned representations are then applied in the reconstruction process to infer missing data samples. This type of reconstruction has been performed in the visualization image domain to infer high-resolution images from given low-resolution images of isosurfaces \citebody{weiss2019isosuperres}, in the spatial domain to infer a higher resolution of a 3D data set from a low-resolution version~\citebody{zhou2017volume}, and in the temporal domain to infer a temporally dense volume sequence from a sparse temporal sequence~\citebody{han2019tsr}.

Others have even proposed neural networks that are trained end-to-end to learn directly the visual data representations instead of the data itself.
Berger~\etAl ~\citebody{berger2019generative} propose a deep image synthesis approach to assist transfer function design, by letting an artificial neural network synthesize new volume rendered images from only a selected viewpoint and a transfer function. He~\etAl ~\citebody{He2020InSituNet} demonstrate that artificial neural networks can even be used to bridge the data entirely, by learning the relationships between the input parameters of a simulation and visualizations of the simulation results. Both approaches do not make any explicit assumptions about the relevance of certain structures in the data, yet the learned relationships between parameters and visual representations are considered in the image generation process. 

\subsection{Contribution}

Our goal is to make a further step towards learning visual representations, by investigating whether a neural network can a) learn the relevance of structures for generating such representations, b) use this knowledge to adaptively sample a visual representation of a volumetric object, and c) reconstruct an accurate image from the sparse set of samples. \changed{Notably, even we can demonstrate for very large volumes and image sizes that adaptive sampling can save rendering time, performance improvement is not our main objective. It is even fair to say that an optimized GPU volume ray-caster can hardly been beaten performance-wise.
Our main objective is to gain an improved understanding of the learning skills of neural networks for generating visual representations in an unsupervised manner, by letting networks learn the relevance of certain structures for obtaining such representations. It can eventually become possible to generate data representations that compactly encode relevant structures in a way they can be used by a neural network to visualize the data. Such insights can further facilitate the use of transfer learning to construct synthetic data sets that contain the structures that are important for successful learning tasks on real data. For viewpoint selection, a network might learn to recommend views showing many important structures, and for training this information can be used to acquire more data from similar views.}

To address our objectives, we introduce a novel network pipeline that is trained end-to-end to learn the relevance of certain structures in the data for generating a visual representation (\autoref{fig:teaser}). This pipeline is comprised of two consecutive internal network stages: An importance network and a reconstruction network. Both networks work in tandem, in that the first learns to place samples along relevant structures by using the second network to give feedback on how well a visual representation of the data can be reconstructed from the sparse sampling. 
Our approach differs from previous adaptive sampling approaches in volume visualization~\citebody{levoy1990volume, kratz2011adaptive,Belyaev2018-adaptiveiso} in that it does not rely on any specific saliency model to determine the image regions that need to be refined. In contrast, we propose a network-based processing pipeline that simultaneously learns where to sample and how to accurately reconstruct an image from the sparse samples, solely using losses on the reconstructed images.

\changed{For learning an importance map from a low-resolution visualization and reconstructing an image from a sparse set of pixel values, we use two modified versions of an EnhanceNet ~\cite{Sajjadi2017Enhance}. To enable network-based learning using gradient descent, two novel processing stages are introduced:
\begin{itemize}
    \item A differentiable sampling stage that models the relationship between sample positions and visual representation.
    \item \changed{A differentiable image reconstruction stage using the pull-push algorithm~\citebody{gortler1996lumigraph,kraus2009pull} to model the relationship between a sparse set of image samples and the reconstructed image.}
\end{itemize}
}

In a number of experiments, we demonstrate that the importance network effectively selects structures that are relevant for the final visual representation. We focus on adaptive sampling in image-space, i.e., using surface samples and samples resulting from direct volume rendering.  As a future direction of research, we outline adaptive sampling in object space, i.e., using data samples along view-rays. Our experiments include qualitative and quantitative evaluations, which indicate good reconstruction accuracy even from few samples. 
The source code of our processing pipeline is available at \url{https://github.com/shamanDevel/AdaptiveSampling}, including some of the data sets that have been used for training and validation.

\section{Related Work}

In the following, we review previous works that share similarities with our approach from the fields of adaptive sampling for rendering as well as neural network-based image and volume reconstruction. 

{\bf Adaptive Sampling for Rendering}
Adaptive rendering has a long tradition in computer graphics, to reduce the number of rays to trace against the scene and perform rasterization at lower image resolution. At the core of such approaches is the computation of importance values to steer the adaptive refinement, for instance, based on perceptual models ~\citebody{Bolin98-perceptuallybasedsampling,Myszkowski1998-VDP,Ramasubramanian1999-perceptually}, image saliency models using pixel variance~\citebody{painter1989antialiased,rigau2003refinement}, image difference operations ~\citebody{Longhurst2008-gpusaliency}, or entropy-based measures~\citebody{xu2005adaptive}, to name just a few. In the context of foveated rendering~\citebody{guenter2012foveated}, where usually a static adaptive sampling pattern is used that moves with the users gaze, a luminance-contrast-aware criterion was introduced to enable feature-aware adaptivity ~\citebody{okan2019luminance}. 
The importance map generation process is often started from an image preview that is calculated using a low resolution render pass or a high-resolution estimate that can be created in a significantly faster way than the final image.

For volume rendering, a number of approaches have investigated adaptive sampling in object space, to reduce the number of samples along the view rays ~\citebody{Novins92-controlledprecision,Danskin92-fastvolren,Lindholm2013TowardsDC,Campagnolo2015-rayadapt}. 
Adaptive image-space refinement has been proposed by Levoy~\citebody{levoy1990volume}, by using the color variances between pixels at low image resolution to decide whether to refine the image resolution locally. Kratz~\etAl~\citebody{kratz2011adaptive} propose to use the difference image between two coarser resolution images, and locally refine where high differences are observed. Belayev ~\etAl~\citebody{Belyaev2018-adaptiveiso} render low-resolution images of isosurfaces and refine depending on how many pixels surrounding a pixel in the low-resolution view fulfill certain requirements. Frey ~\etAl~\citebody{Frey2014-progressive} use a fixed random sampling structures that is applied in a hierarchical manner to progressively refine the image.

The major differences between these approaches and our proposed sampling pipeline are as follows: Firstly, the pipeline learns to adapt the sampling in an unsupervised manner. A specific feature descriptor that steers the placement of samples is not used, and importance values are learned solely using losses on the reconstructed image. Secondly, the number of samples can be prescribed, which is not easily possible with existing schemes due to their pixel-iterative nature. Thirdly, the pipeline learns simultaneously the adaptive sampling and the image reconstruction from the sparse set of samples. In all previous schemes, the final interpolation step is decoupled from the sampling process.

{\bf Deep Learning for Upscaling and Denoising}
In recent years, deep learning approaches have been used successfully for single-image and video super-resolution tasks~\citebody{dong2014learning, shi2016real, tai2017image, tao2017detail,sajjadi2018frame,chu2018temporally}, i.e., the upscaling of images and videos from a lower to some higher resolution. 
Many previous works let the networks learn to optimize for losses between the inferred and ground-truth images based on direct vector norms~\citebody{kim2016deeply,kim2016accurate}. GANs were introduced to prevent the undesirable smoothing of direct loss formulations~\citebody{Sajjadi2017Enhance,ledig2017photo}, and instead use a second network that discriminates real from generated samples and guides the generator. Convolutional architectures~\citebody{dong2014learning} with residual blocks~\citebody{he2016deep} are popular generator architectures that offer training stability as well as high-quality inference. Losses based on the feature-space differences of image classification networks, e.g., a pre-trained VGG network~\citebody{johnson2016perceptual}, have shown to mimic well the human's capability to assess the perceptual similarity between two images. 

\changed{The approach closest to ours is by Kuznetsov~\etAl~\citebody{kuznetsov2018deep} for learning adaptivity in Monte-Carlo path-tracing and denoising of the final image. A first network learns to adapt the number of additional paths from an initial image at the target resolution, which is generated via one path per pixel. A second denoising network learns to model the relationship between an image with increased variance in the color samples to the ground truth rendering ~\citebody{cnn-denoise,mara17towards}. Conceptually, our approach differs in that it works on a low-resolution input map and then learns to freely position the sample locations in image space, i.e. it learns to place zero or one sample per pixel. This requires a completely different differentiable sampling stage, as well as a differentiable image reconstruction stage that can work on a sparse set of samples. 
Furthermore, Kuznetsov~\etAl{} use finite differences between images of different sample counts for gradient estimation. Incurring noise is reduced by averaging multiple samples with different sample counts, which is not possible in our approach where at most one sample per pixel is taken. Instead, we propose a sigmoid approximation that can be differentiated analytically.}

In visualization, Zhou~\etAl~\citebody{zhou2017volume} presented a CNN-based solution that upscales a volumetric data set using three hidden layers designed for feature extraction, non-linear mapping, and reconstruction, respectively. 
Han~\etAl~\citebody{han2019flow} introduced a two-stage approach for vector field reconstruction via deep learning, by refining a low-resolution vector field from a set of streamlines.
Berger~\etAl~\citebody{berger2019generative} proposed a deep image synthesis approach to assist transfer function design using GANs, by letting a network synthesis new volume rendered images from only a selected viewpoint and a transfer function.  
The use of neural network-based inference of data samples in the context of in situ visualization was demonstrated by Han and Wang~\citebody{han2019tsr}, where a network learns to infer missing time steps between 3D simulation results. 
He~\etAl~\citebody{He2020InSituNet} use neural networks for parameter-space exploration, by training a network to learn the dependencies between visual mappings of simulation results and the input parameters of the simulation. 
Guo~\etAl~\citebody{guo2020ssrvfd} designed a deep learning framework that produces coherent spatial super-resolution of 3D vector field data. 
Weiss~\etAl~\citebody{weiss2019isosuperres} extent image upscaling to geometry images of isosurfaces including depth and normal information. Instead of data upscaling, Tkachev~\etAl~\citebody{tkachev2019prediction} predict a next time-step of a simulation and identify regions of interest by high variance between the network prediction and the ground truth. Common to all these approaches is the use of a regular sampling structure that does not consider the importance of samples in the inference step.

\section{Learning to Sample}\label{sec:method}

\begin{figure*}%
\centering%
\resizebox{0.95\linewidth}{!}{\input{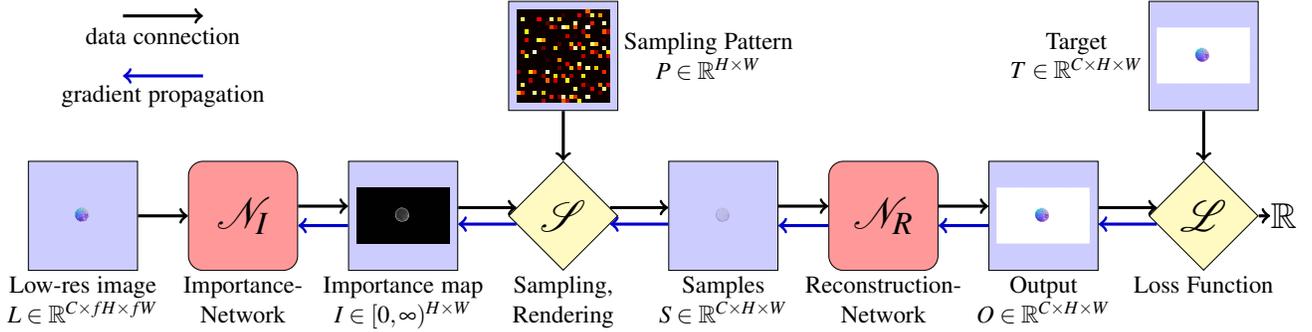}}%
\caption{Overview of network-based adaptive sampling. From a low-resolution image $L$, the importance network infers the importance map $I$. The sampler $\mathcal{S}$ uses this map together with a sampling pattern $P$ to adaptively place samples in the high-resolution image $S$. Ray-casting the object at these samples generates a sparse image. The reconstruction network recovers the dense output $O$.}
\label{fig:pipeline}
\end{figure*}

In the following, we discuss how the importance network makes use of both the adaptive sampling stage and the reconstruction network to learn where to place samples with higher density. The importance network (\autoref{sec:method:importance}) receives an image of the data set at low resolution. This image $L$ is of shape $C \times fH \times fW$, where $W$ and $H$ denote the screen resolution, and $f$ the downsampling factor. This factor is set to $1/8$ in all of our experiments. Each image pixel is comprised of $C$ channels, such as color, depth, and normal, representing what is seen through that pixel. The network is trained to learn an importance function $\mathcal{N}_I$ that generates a gray-scale importance map $I\in[0,\infty)^{H\times W}$ in which low and high values, respectively, indicate where less or more samples are taken.

The sampler $\mathcal{S}$ (\autoref{sec:method:sampling}) takes the importance map and places a given number of samples, e.g., $5\%$ of the pixel, in the full resolution image $S\in\mathbb{R}^{C \times H \times W}$ according to the importance information. Only at these samples the object is rendered.
The reconstruction network learns a function $\mathcal{N}_R$ (\autoref{sec:method:reconstruction}) that reconstructs the final output $O\in\mathbb{R}^{C \times H \times W}$ from the sparse set of samples.
We make the sampler differentiable w.r.t. sample positions to allow gradient flow from the reconstruction network (\autoref{sec:method:reconstruction}) to the importance network, so that the reconstruction network is trained simultaneously and propagates the loss information to the sampling stage. 
Since the entire pipeline is trained end-to-end using a loss on the reconstructed and ground truth images, the importance network and the pair of sampler and reconstruction network work together in an effort to learn the placement of samples so that high reconstruction quality is achieved. 

In principle, one can refrain from using \changed{a separate importance map}, by realizing the sampler as a network that directly learns the adaptive sampling. In this case, however, modelling the positional information in a network requires to represent positions explicitly, either in a graph structure or a linear field, so that less efficient graph networks or fully-connected networks need to be used. 
Furthermore, the sampler has to be re-trained whenever a different number of samples is used. Our approach enables to use efficient convolutional networks, and to change the number of samples at testing time.

An overview of the processing pipeline is shown in \autoref{fig:pipeline}. It works with images comprised of an arbitrary number $C$ of channels.  
In the first part of this work, the pipeline is introduced for isosurface rendering with $C=5$, i.e., a binary mask (1: hit, 0: no hit), a normal vector, and a depth value.
The application to direct volume rendered images is discussed in \autoref{sec:dvr}.

\changed{Weiss~\etAl~\citebody{weiss2019isosuperres} enforce frame-to-frame coherence during animations by including a temporal loss in the training step. This loss considers the difference between the previous frame -- warped by the frame-to-frame optical flow -- to the current frame.
In the accompanying video, this approach is used for both the importance and reconstruction network. In the following discussion, however, temporal connections are omitted and the focus is solely on single image reconstruction for clarity.
}

\subsection{Importance Network}\label{sec:method:importance}

The importance network $I\leftarrow\mathcal{N}_I(L)$ determines the distribution of the samples that are required by the reconstruction network to generate the visual output according to some loss function.
Deeming every pixel equally important, i.e., 
\begin{equation}
	\mathcal{N}_{I,\text{constant}}(L)_{ij} = \mathbf{1},
\label{eq:importanceConstant}
\end{equation} 
leads to a uniform distribution of the samples~\citebody{zhou2017volume, han2019tsr,weiss2019isosuperres}.
Alternatively, and in the spirit of classical edge detection filters, the screen space gradients of the individual channels can be used, i.e., 
\begin{equation}
	\mathcal{N}_{I,\text{gradient}}(L)_{ij} = \sum_c{w_c ||\nabla L_{ij,c}||_2^2},
\label{eq:importanceGradient}
\end{equation}
where $\nabla L_{ij,c}$ is the screen space gradient of channel $c$ at location $ij$. The contributions of the individual channels are weighted by $w\in\mathbb{R}^C$.
Other known importance measures consider screen space curvature via the variation of surface normals~\citebody{prantl2016fast}, or color contrast via the variation of luminance~\citebody{okan2019luminance}.

Alternatively, we introduce a fully convolutional neural network $\mathcal{N}_{I,\text{net}}$ (\autoref{sec:training:architectures}) that predicts a high-resolution greyscale importance map $I$ from a low-resolution rendering $L$. 
Notably, this network is not trained w.r.t. specific characteristics that are derived from the image like gradients or luminance information, since this requires to heuristically decide on the importance of pixels. Instead, it is trained end-to-end with losses only on the reconstructed color information, by gradient descend all along the processing pipeline. In \autoref{sec:results}, network-based inference of the importance map is compared to alternative approaches, showing superior prediction of regions that are important for the final image.

\subsection{Differentiable Sampling}\label{sec:method:sampling}

Given the target number of samples in the final image, e.g. $\mu=5\%$ of all pixels, the sampler uses the importance map $I$ to determine where to place these samples.
To generate the given number of samples, two main classes of algorithm are commonly used in rendering:
\begin{itemize}
	\item Stippling starts with a given number of points at random locations and iteratively optimize these locations so that the point density matches the density of the importance map \citebody{deussen2017stippling,gortler2019stippling}.
	\item Importance sampling treats the importance map as a density function and place samples via rejection sampling or the inverse cumulative distribution function \citebody{lawrence2005adaptive,bashford2013importance}.
\end{itemize}
These algorithms, however, are not easily differentiable w.r.t. changes in the importance map, since they use discrete optimizations or random processes, and often are too slow for real-time applications.
To make the sampling process differentiable and fast, we propose a sampling strategy that computes for every pixel independently the chance of being sampled. This is achieved by a smooth approximation of rejection sampling, which is differentiable and allows for gradient propagation through the network pipeline. Since every pixel can be processed independently, this scheme can effectively leverage parallel execution on the GPU. On the other hand, it does not allow for an exact match of the prescribed number of samples, yet produces a number of samples that slightly varies around the target number.

In a first step, the importance map $I$ is normalized to have a prescribed mean $\mu$ and minimal value $l\leq\mu$.
Let $\mu_I$ be the mean of $I$ over all pixels, then the image
\begin{equation}
	I'_{ij} := \min\left\{1, l + I_{ij}\frac{\mu-l}{\mu_I+\epsilon}\right\}
\label{eq:samplingNormalization}
\end{equation}
has the desired properties. A small constant $\epsilon=10^{-7}$ is used to avoid division by zero. 
The minimal value $l$ is required to maintain a lower bound on the sample distribution in empty areas, which is important to allow for an accurate reconstruction in such areas. We use $l=0.002$ in all of our experiments.
Clamping to a maximal value of 1 is required by the following sampling step, which is realized as an independent Bernoulli process via rejection sampling, i.e., 
a sample at location $ij$ is taken if the probability $I'_{ij}$ is larger than a uniform random value $x\in[0,1]$.

To make the sampling deterministic and parallelizable on the GPU, a sampling pattern $P\in[0,1]^{H\times W}$ -- uniformly distributed in $[0,1]$ -- is first generated by using a permutation of the numbers $\frac{1}{HW}\{0,...,HW-1\}$.
We analyze four different strategies for generating the permutations: Random sampling, regular sampling, Halton sampling~\citebody{halton1964sampling}, and plastic sampling~\citebody{roberts2020plastic}.
Plastic sampling has been selected, since it produced slightly superior results in all of our experiments. \autoref{app:sampling} provides a detailed evaluation of the different strategies.

Ray-casting is then used to compute what is seen through the pixels at the determined sample locations. This information is stored in the high resolution image $S\in\mathbb{R}^{C \times H \times W}$. Since during training the same view is rendered many times using different sampling patterns, pre-computed high-resolution target images $T\in\mathbb{R}^{C \times H \times W}$ are provided with the low-resolution inputs. 
Then, the sampling process simply becomes a selection of pixels from $T$:
\begin{equation}
	S_{ij} = \mathbbm{1}_{I'_{ij}-P_{ij}} T_{ij}, 
\label{eq:samplingSelection}
\end{equation}
where $\mathbbm{1}_x$ is $1$ if $x>0$ else 0.
Since the sampling function in \autoref{eq:samplingSelection} is a step function with zero gradients almost everywhere, it is not differentiable w.r.t. the importance map $I$, from which $I'$ is derived. Correspondingly, gradients in the loss function w.r.t. the weights and biases of the importance network will  also be zero. Therefore, \autoref{eq:samplingSelection} is approximated with a smooth sigmoid function to make it differentiable, so that gradients of the loss function can be back-propagated through all network stages to change the importance map accordingly. Then, the sampling function becomes
\begin{equation}
	S_{ij} = \operatorname{sig}\left(\alpha (I'_{ij}-P_{ij})\right) T_{ij}, \ \ \
	\text{sig}(x) := \frac{1}{1+e^{-x}},
\label{eq:samplingSigmoid}
\end{equation}
where $\alpha>0$ determines the steepness of the function. 
The differentiable approximation is used only in the training phase, while in the validation phase the ray-caster renders the discrete samples obtained via rejection sampling.
For $\alpha\rightarrow\infty$, \autoref{eq:samplingSigmoid} converges to \autoref{eq:samplingSelection}. A large value of $\alpha$ leads to samples that are either very close to 0 or 1, but leads to exploding gradients in the backward pass. A low value leads to samples that smoothly cover the entire interval between 0 and 1. In this case, however, the mismatch between the ``fractional'' samples that are used only during training and the discrete ``binary'' samples that are used for testing and validation leads to a significant reduction of the reconstruction quality.
In our experiments, a value around $\alpha=50$ always lead to the best results. Going beyond 50 quickly introduces floating-point precision issues and exploding gradients thereof. An evaluation of the dependency between the value of $\alpha$ and the reconstruction quality is provided in \autoref{sec:results:validation}.

\subsection{Differentiable Reconstruction}\label{sec:method:reconstruction}

Given the sparse set of samples $S$, the reconstruction function $\mathcal{N}_R$ needs to estimate the undefined pixel values to produce the dense high-resolution output image $O\in\mathbb{R}^{C\times H\times W}$.
By using a differentiable reconstruction function, gradients of the loss function on the reconstructed images and the ground truth image can be back-propagated through the sampling stage to the importance map. 

In principle, there are different possibilities to fulfill the requirement of differentiability: Firstly, a neural inpainting network can be trained on sparse inputs and the ground truth outputs to learn the reconstruction. However, as we have verified in a number of experiments, network-based inpainting \citebody{iizuka2017globally,Liu2018PartialConv,yu2019free} at a sparsity level as used in our application leads to low reconstruction quality (see \autoref{fig:results:reconstruction}b). The highly varying sample density with gaps between valid pixel values of up to 20 pixels poses a challenging problem for known network architectures.
Furthermore, since during training the sampling mask in our proposed pipeline is not binary but contains continuous values, techniques like Partial Convolutions~\citebody{Liu2018PartialConv} are not applicable.

Secondly, classical non-network-based inpainting methods can be employed, for instance, PDE-based methods solving a constrained Laplace problem  \citebody{Bertalmio2001NavierInpainting,Getreuer2012TVInpainting}, or patch-based methods using non-local cost functions involving correspondence functions \citebody{Igehy1997PatchInpainting,Efros1999PatchInpainting,Criminisi2004Inpainting}. 
These methods, however, are not easily differentiable w.r.t. the sampling mask. For example, PDE-based methods use the samples as Dirichlet boundaries and, to the best of our knowledge, there is no meaningful interpretation of a ``fractional'' Dirichlet boundary. Patch-based methods, on the other hand, use a discrete search over the image space to find a correspondence function, which makes the derivation of continuous gradients impossible.

Therefore, we introduce a novel reconstruction approach that combines a differentiable inpainting method with a residual neural network that learns to improve the inpainting result. 
In particular, we propose a variation of the pull-push algorithm~\citebody{gortler1996lumigraph,kraus2009pull}, which is differentiable with respect to the sampling mask and can cope with a mask that comprises fractional values.

\begin{figure}[t]
\centering
\input{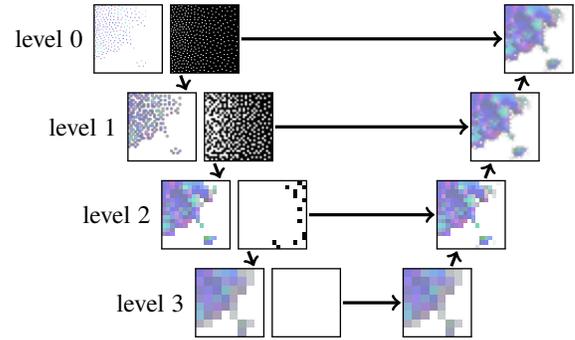}
\caption{Pull-push-based inpainting using a mipmap hierarchy of image samples and masks. The image is downsampled until all pixels are filled, and then upsampled by combining interpolated values from lower levels with the pixels at the current level. Masks are propagated through the hierarchy to obtain proper interpolation weights.}%
\label{fig:inpainting1}%
\end{figure}

The pull-push algorithm builds upon the idea of mipmap hierarchies.
Firstly, the sparsely sampled high-resolution image and the mask are recursively filtered and downsampled by a factor of $2$. The pixel values are averaged using the fractional values in the sampling mask as weights (average pooling), and max-pooling is used to combine the values in the mask. This has the effect of filling the undefined pixels with values that are averaged from a gradually increasing surrounding. Upon reaching a termination criterion, either a maximal number of steps or complete restoration of the undefined pixels, the images are bilinearly upscaled again. During upscaling, the pixel values from the coarse levels are weighted by the values in the mask at this level, and they are then blended with the value at the fine level based on the sampling values at that level. 
This allows to smoothly transition from filled pixels at the fine level that are kept in the output towards interpolated values for lower values in the sampling mask.
A schematic illustration of the process is shown in \autoref{fig:inpainting1}.
Since the algorithm makes use exclusively of continuous pooling and interpolation operations, it is fully differentiable with respect to changes in the pixel data and the sampling mask.
The forward code and a manually derived backward code are given in \autoref{app:pullpush}. The algorithm has been implemented via custom CUDA operations in PyTorch~\citebody{PyTorch2019}.

After inpainting the sparse samples via the pull-push algorithm, a fully convolutional network is used to improve the reconstruction by modeling the relationship between the inpainting result and the ground truth. The network sharpens the results and resolves blurred silhouettes created by the inpainting algorithm.
We use the EnhanceNet~\citebody{Sajjadi2017Enhance} as base architecture for this learning task, which is discussed in detail in \autoref{sec:training:architectures}. 
In particular, we use the EnhanceNet as a residual network that starts with the inpainting result and learns to infer the changes to the reconstructed samples. A quantitative comparison of different learning approaches is provided in \autoref{sec:results:validation}.

\section{Training Methodology}\label{sec:training}

In this chapter, we provide a detailed discussion of the used
network architectures, as well as the training and inference steps.
We also shed light on the dependency of the reconstruction quality
on the used loss functions.

\subsection{Training Data}\label{sec:training:data}

As training and validation input, 5000 images of randomly selected isosurfaces in the Ejecta data set, a particle-based supernova simulation, were generated via GPU ray-casting at a screen resolution of $512^2$. Each time step was resampled to Cartesian grids with a resolution of $256^3$ and $512^3$. 
The surfaces were rendered from random camera positions, at varying distance to the object and always facing the object center. Renderings are taken from different time steps and resolution levels to let the pipeline learn features at different granularity~\citebody{weiss2019isosuperres}.
The renderer provides the normals at the surface points, which are used in a post-process to compute colors via the Phong illumination model.
From this image set, about 20.000 random crops of size $256^2$ and showing the isosurface in at least $50\%$ of the pixels were taken, and split between training ($80\%$) and validation ($20\%$).
For training, the mean importance value was set to $\mu=0.1$, i.e., $10\%$ of the samples (see \autoref{eq:samplingNormalization}). This does not prohibit using less samples for validation and testing, yet we found it beneficial to allow the network to use more samples during training. We used the Adam~\citebody{kingma2014adam} optimizer with a learning rate of $10^{-4}$. 
The networks were trained on a single GeForce GTX 1080 for 300 to 500 epochs in around 5-6 days.

\subsection{Network Architectures}\label{sec:training:architectures}

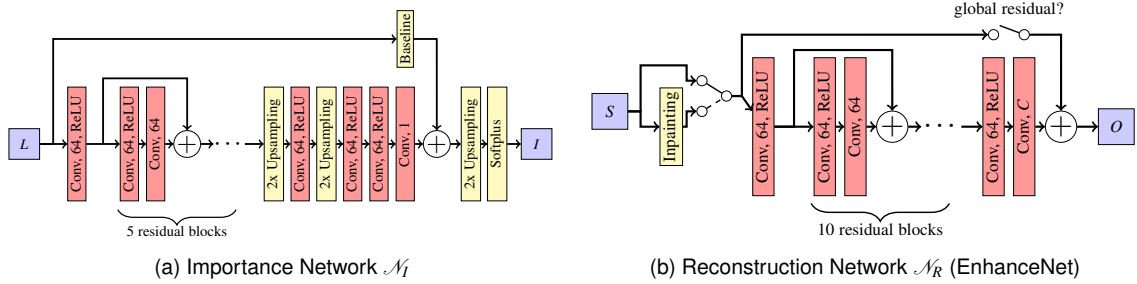
\begin{figure*}%
\centering
\subfloat[Importance Network $\mathcal{N}_I$\label{fig:architectures:imp}]{\resizebox{0.4\linewidth}{!}{\begin{tikzpicture}[scale=0.6,
	every text node part/.style={align=center},
	every node/.style={minimum size=0.7cm,inner sep=0pt},
	every label/.style={shape=rectangle, draw=none, fill=none, minimum size=0pt,inner sep=2pt},
	data/.style={draw,rectangle,fill=blue!20},
	op/.style={draw,rectangle,inner sep=2pt,fill=yellow!30,rotate=90,minimum size=0cm,minimum width=2.6cm},
	base/.style={draw,rectangle,inner sep=2pt,fill=yellow!30,rotate=90,minimum size=0cm},
	net/.style={draw,rectangle,inner sep=2pt,fill=red!40,rotate=90,minimum size=0cm,minimum width=2.6cm},
	dataflow/.style={very thick,->},
	tmpflow/.style={very thick,->,yellow!40!black,dashed}]
	
\node[data] (I1) at (0,0.0) {$L$};

\node[net] (C1) at (2,0) {Conv, 64, ReLU};

\node[net] (C2) at (4,0) {Conv, 64, ReLU};
\node[net] (C3) at (5,0) {Conv, 64};
\node[minimum size=0,circle,draw,inner sep=0pt] (A1) at (6.2,0) {\LARGE $+$};
\node[minimum size=0] (dot1) at (7.8,0) {\LARGE $\cdots$};
\draw [decorate,decoration={brace,amplitude=10pt},xshift=-4pt,yshift=0pt,thick]
	(8.2,-2.5) -- (3.7,-2.5) node [black,midway,yshift=-0.5cm] 
	{5 residual blocks};

\node[op] (O1) at (9.5,0) {2x Upsampling};
\node[net] (C4) at (10.5,0) {Conv, 64, ReLU};
\node[op] (O2) at (11.5,0) {2x Upsampling};
\node[net] (C5) at (12.5,0) {Conv, 64, ReLU};
\node[net] (C6) at (13.5,0) {Conv, 64, ReLU};
\node[net] (C7) at (14.5,0) {Conv, 1};

\node[base] (Base) at (14.5,4) {Baseline};
\node[minimum size=0,circle,draw,inner sep=0pt] (A2) at (15.7,0) {\LARGE $+$};
\node[op] (O3) at (17,0) {2x Upsampling};
\node[op] (O4) at (18,0) {Softplus};

\node[data] (Out) at (19.5,0) {$I$};

\draw[dataflow] (I1) -- (1.1,0.0) -- (1.1, 0) -- (C1);
\draw[dataflow] (I1) -- (1.1,0.0) -- (1.1, 4) -- (Base);
\draw[dataflow] (C1) -- (C2);
\draw[dataflow] (C2) -- (C3);
\draw[dataflow] (C3) -- (A1);
\draw[dataflow] (A1) -- (dot1);
\draw[dataflow] (dot1) -- (O1);
\draw[dataflow] (3,0) -- (3,2.5) -- (6.2,2.5) -- (A1);
\draw[dataflow] (O1) -- (C4);
\draw[dataflow] (C4) -- (O2);
\draw[dataflow] (O2) -- (C5);
\draw[dataflow] (C5) -- (C6);
\draw[dataflow] (C6) -- (C7);
\draw[dataflow] (C7) -- (A2);
\draw[dataflow] (Base) -- (15.7,4) -- (A2);
\draw[dataflow] (A2) -- (O3);
\draw[dataflow] (O3) -- (O4);
\draw[dataflow] (O4) -- (Out);

\end{tikzpicture}}}%
~~~~~%
\subfloat[Reconstruction Network $\mathcal{N}_R$ (EnhanceNet)\label{fig:architectures:recEnhance}]{\resizebox{0.4\linewidth}{!}{\begin{tikzpicture}[scale=0.6,
	every text node part/.style={align=center},
	every node/.style={minimum size=0.7cm,inner sep=1pt},
	every label/.style={shape=rectangle, draw=none, fill=none, minimum size=0pt,inner sep=2pt},
	data/.style={draw,rectangle,fill=blue!20},
	op/.style={draw,rectangle,inner sep=2pt,fill=yellow!30,rotate=90,minimum size=0cm,minimum width=2.7cm},
	base/.style={draw,rectangle,inner sep=2pt,fill=yellow!30,rotate=90,minimum size=0cm},
	net/.style={draw,rectangle,inner sep=2pt,fill=red!40,rotate=90,minimum size=0cm,minimum width=2.7cm},
	dataflow/.style={very thick,->},
	tmpflow/.style={very thick,->,yellow!40!black,dashed}]
	
\node[data] (I1) at (-3,0.5) {$S$};

\node[op,minimum width=0cm] (Imp) at (-1,0) {Inpainting};

\node[draw,circle,fill=none,minimum size=0.2cm] (S1) at (0,0.5) {};
\node[draw,circle,fill=none,minimum size=0.2cm] (S3) at (0.8,1) {};
\node[draw,circle,fill=none,minimum size=0.2cm] (S2) at (0,1.5) {};

\node[net] (C1) at (2,0) {Conv, 64, ReLU};

\node[net] (C2) at (4,0) {Conv, 64, ReLU};
\node[net] (C3) at (5,0) {Conv, 64};
\node[minimum size=0,circle,draw,inner sep=0pt] (A1) at (6.2,0) {\LARGE $+$};
\node[minimum size=0] (dot1) at (7.8,0) {\LARGE $\cdots$};
\draw [decorate,decoration={brace,amplitude=10pt},xshift=-4pt,yshift=0pt,thick]
	(8.2,-2.5) -- (3.7,-2.5) node [black,midway,yshift=-0.5cm] 
	{10 residual blocks};

\node[net] (C6) at (9.5,0) {Conv, 64, ReLU};
\node[net] (C7) at (10.5,0) {Conv, $C$};

\node[draw,circle,fill=none,minimum size=0.2cm] (S4) at (9.5,3) {};
\node[draw,circle,fill=none,minimum size=0.2cm] (S5) at (10.5,3) {};
\node[minimum size=0,circle,draw,inner sep=0pt] (A2) at (11.7,0) {\LARGE $+$};

\node[data] (Out) at (13.5,0) {$O$};

\draw[dataflow] (I1) -- (-2,0.5) -- (-2,0) -- (Imp);
\draw[dataflow] (Imp) -- (-0.5,0) -- (-0.5,0.5) -- (S1);
\draw[dataflow] (I1) -- (-2,0.5) -- (-2,2) -- (-0.5,2) -- (-0.5,1.5) -- (S2);
\draw[thick] (S2) -- (S3);
\draw[thick,dashed] (S1) -- (S3);
\draw[dataflow] (S3) -- (1.3,1) -- (C1);

\draw[dataflow] (C1) -- (C2);
\draw[dataflow] (C2) -- (C3);
\draw[dataflow] (C3) -- (A1);
\draw[dataflow] (C1) -- (3,0) -- (3,2.5) -- (6.2,2.5) -- (A1);
\draw[dataflow] (A1) -- (dot1);
\draw[dataflow] (dot1) -- (C6);
\draw[dataflow] (C6) -- (C7);
\draw[dataflow] (C7) -- (A2);
\draw[dataflow] (S5) -- (11.7,3) -- (A2);
\draw[dataflow] (S3) -- (1.3,1) -- (1.3,3) -- (S4);
\draw[dataflow] (A2) -- (Out);

\draw[thick] (9.7,3.3) -- (S5);
\node[] at (10,3.8) {global residual?};

\end{tikzpicture}}}%
\caption[Network architectures used in the proposed pipeline.]{Network architectures used in the proposed pipeline: To estimate the importance map, we use a smaller version of the EnhanceNet~\citebody{Sajjadi2017Enhance} with a 4x-upsampling factor, a residual connection with screen space gradient magnitude as baseline, and a 2x-upsampling network as a post-process. For the reconstruction network, we experimented with the option of passing the raw samples or interpolated samples as input and using a global residual connection or not. As network architecture, an EnhanceNet with 10 residual blocks is used.}%
\label{fig:architectures}%
\end{figure*}

The proposed sampling pipeline comprises two trainable blocks: The importance network $\mathcal{N}_I$ and the reconstruction network $\mathcal{N}_R$. Both networks use 3x3 convolutions with zero-padding and a stride of one. 
The importance network is a variant of EnhanceNet~\citebody{Sajjadi2017Enhance}, yet with only 5 residual blocks (\autoref{fig:architectures:imp}). Instead of directly estimating the importance map, the network takes as input an importance map that is computed using screen space gradient magnitudes (\autoref{sec:method:importance}), and learns to improve this map using a residual connection. We refer to \autoref{sec:results:validation} for a quantitative comparison of the network results w/ and w/o an initial gradient-based importance estimate.

The importance network performs 4x-upscaling of a low-resolution input image with $1/4$ the resolution of the final image. Thus, generating the input image requires to sample $1/4^2=6.25\%$ of the pixels in the target image, which already exceeds a prescribed limit of, e.g., $5\%$ of the pixels. Therefore, an image with $1/8$ the final resolution is used as input, and the network performs 4x-upscaling to an intermediate image with $1/2$ the final resolution, followed by an additional 2x-upscaling of the inferred importance map. This allows to more aggressively reduce the number of initially required samples, i.e., only $1/8^2\approx 1.56\%$ of the pixels in the final importance map need to be rendered. 

The reconstruction network $\mathcal{N}_R$ estimates the mask, normal, and depth values at all pixels, thereby also changing the initial values that were drawn in the rendering process. A modified EnhanceNet (\autoref{fig:architectures:recEnhance}) shows superior reconstruction results compared to alternative architectures such as the U-Net~\citebody{ronneberger2015UNet}. Let us refer to \autoref{app:unet} for a more detailed analysis of both architectures. Both networks are provided in the code repository accompanying this paper. 

Our experiments (\autoref{sec:results:validation}) show improved reconstruction quality if inpaining is performed first and the result is then passed to a network that uses a residual connection to learn the differences between this result and the ground truth.
In addition to the inpainted input samples, we pass the sample mask to the network as a per-sample measure of certainty.
Since the network produces output values in $\R$, both the mask and depth values are clamped to $[0,1]$ and the normals are scaled to unit length before shading is applied.

\subsection{Loss Functions}\label{sec:training:losses}
We employ regular vector norms between the network prediction $O$ and the target image $T$ as primary loss functions on the individual output channels.
Since the $L_2$ norm tends to smooth out the resulting images, we make use of the $L_1$ norm in this work. With the channels of the output image, i.e., the mask $M$, the normal map $N$, and the depth $D$, given as subscript, the $L_1$ loss of a selected channel $X$ is 
\begin{equation}
    \mathcal{L}_{1,X}=||T_{X}-O_{X}||_1 .
    \label{eq:loss:pixel}
\end{equation}
We do not employ additional perceptual losses, which were shown less effective for isosurface upsampling tasks~\citebody{weiss2019isosuperres}.

The mask channel has a special meaning as it indicates whether or not a ray hits the isosurface. It is used in the final output to perform a hard selection between the reconstructed color values and the background. To make the mask differentiable, however, its values must be continuous, leading to a smooth blend rather than a binary decision.
While this is acceptable along the silhouettes, in the interior it would noticeably distort the reconstruction.
\changed{In principle, via a sigmoidal mapping it can be enforced that the mask values spread continuously between 0 and 1, yet we observed undesirable blurring when using this approach. To produce sharp masks that are either close to zero or one, we therefore constrain the reconstruction via two losses that are added to the regular $L_1$ loss on the mask.}
The first loss is a binary cross entropy (BCE) loss that ''pulls`` the values closer to either zero or one than a normal $L_1$ loss:
\begin{equation}
    \mathcal{L}_{\text{bce}} = -\frac{1}{WH}\sum_{ij}\left(T_{M,ij} \log(O_{M,ij}) + (1-T_{M,ij}) \log(1-O_{M,ij})\right) .
\end{equation}
The BCE loss, however, requires that the output mask lies within $[0,1]$ and thus the mask is clamped beforehand. This leads to zero gradients once the mask reaches values outside of $[0,1]$. Therefore, we add the loss
\begin{equation}
    \mathcal{L}_{\text{bounds}} = \frac{1}{WH}\sum_{ij}\left( \operatorname{max}(0, (2O_{M,ij}-1)^2-1) \right),
\end{equation}
which pushes values outside $[0,1]$ back into $[0,1]$ and leaves values within $[0,1]$ unchanged.

An additional loss term is required to account for the normalization step in \autoref{eq:samplingNormalization}.
The output of the importance map is normalized to limit the number of available samples. Hence, scaling the network output does not influence the values after normalization. Therefore, during training it can happen that the output values increase or decrease in an unbounded manner. To prevent this, a prior on the importance map is used to enforce that the mean is equal to one before the normalization step:
\begin{equation}
    \mathcal{L}_{I,\text{prior}}=\left(1-\frac{1}{WH}\sum_{ij}{I_{ij}}\right)^2 .
    \label{eq:loss:prior}
\end{equation}

The final loss function is a weighted sum of the individual loss terms over all channels, i.e., with $X\in\{M, N, D\}$ it becomes
\begin{equation}
    \begin{aligned}
    \mathcal{L} &= \sum_X{\lambda_X \mathcal{L}_{1,X}} 
                + \lambda_{\text{bce}} \mathcal{L}_{\text{bce}} + \lambda_{\text{bounds}} \mathcal{L}_{\text{bounds}} 
                + \rho\mathcal{L}_{I,\text{prior}}. \\
    \end{aligned}
\end{equation}
Loss weights around
$\lambda_M=5, \lambda_{\text{bce}}=5, \lambda_{\text{bounds}}=0.01, \lambda_N=50, \lambda_D=5$, and $\rho=0.1$ lead to equally good reconstruction quality, while
deviations from these values quickly worsen the reconstruction quality significantly.

\section{Results and Evaluation}\label{sec:results}

In the following, we evaluate the proposed network pipeline.
First, we introduce the quality metrics that are used to compare the results. We then analyze how our design decisions 
influence the reconstruction quality on the validation data (\autoref{sec:results:validation}). These statistics help to identify the network configurations with the best predictive skills. 
Next, the proposed network pipeline is compared to a fixed super-resolution network ( \autoref{sec:results:sr}).
Finally, we shed light on the generalizability of the network pipeline to new views of Ejecta and data sets that were never seen during training (\autoref{sec:results:otherdata}).

\subsection{Quality Metrics}

The quality of network-based reconstruction is assessed using three different image quality metrics commonly used in image processing. These metrics compare the output $O$ of the network pipeline with a ground truth rendering $T$ at the target resolution.

The peak signal-to-noise ratio (PSNR) is based on the $L_2$ loss and is defined as
\begin{equation}
    \text{PSNR}(O, T) = -10 \log_{10}(||O-T||_2^2),
\end{equation}
where $O$ and $T$ are the network output and target image, respectively. 

The Structural Similarity Index (SSIM) \citebody{wang2004image} extends on the idea of per-pixel losses by measuring the perceived quality using the mean and variance of contiguous pixel blocks in the images. It is defined as
\begin{equation}
    \text{SSIM}(O, T) = \frac{(2\mu_O\mu_T+c_1)(2\sigma_{O,T}+c_2)}{(\mu_O^2+\mu_T^2+c_1)(\sigma_O^2+\sigma_T^2+c_2)} ,
\end{equation}
where $\mu_O$ and $\mu_T$ are the average values of $O$ and $T$, $\sigma_O^2$ and $\sigma_T^2$ are the variances of $O$ and $T$, $\sigma_{O, T}$ is the covariance between $O$ and $T$, and $c_1$ and $c_2$ are small constants to avoid division by zero.

We also use the network-based Learned Perceptual Image Patch Similarity (LPIPS) metric \citebody{zhang2018perceptual} that predicts human perception of relative image similarities. LPIPS builds upon a network that is pre-trained on an image classification task---using the AlexNet~\citebody{krizhevsky2014one}---and computes a weighted average of the activations at hidden layers for a given output and target imag.
Note that a lower LPIPS score is better, whereas PSNR and SSIM indicate higher quality by a higher score. Therefore, $1-\text{LPIPS}$ is shown in our statistics for better  comparison.

\subsection{Validation of Design Decisions}\label{sec:results:validation}

Unless otherwise mentioned, all statistics presented in this section were computed on a validation data set using $2000$ novel views of Ejecta at the resolution of $512^2$. The importance map was normalized to have a minimal value $l=0.002$ and a mean value $\mu=0.05$. ''plastic`` sampling was used in the sampling stage.

\begin{figure}
    \centering
    \includegraphics[width=0.4\linewidth]{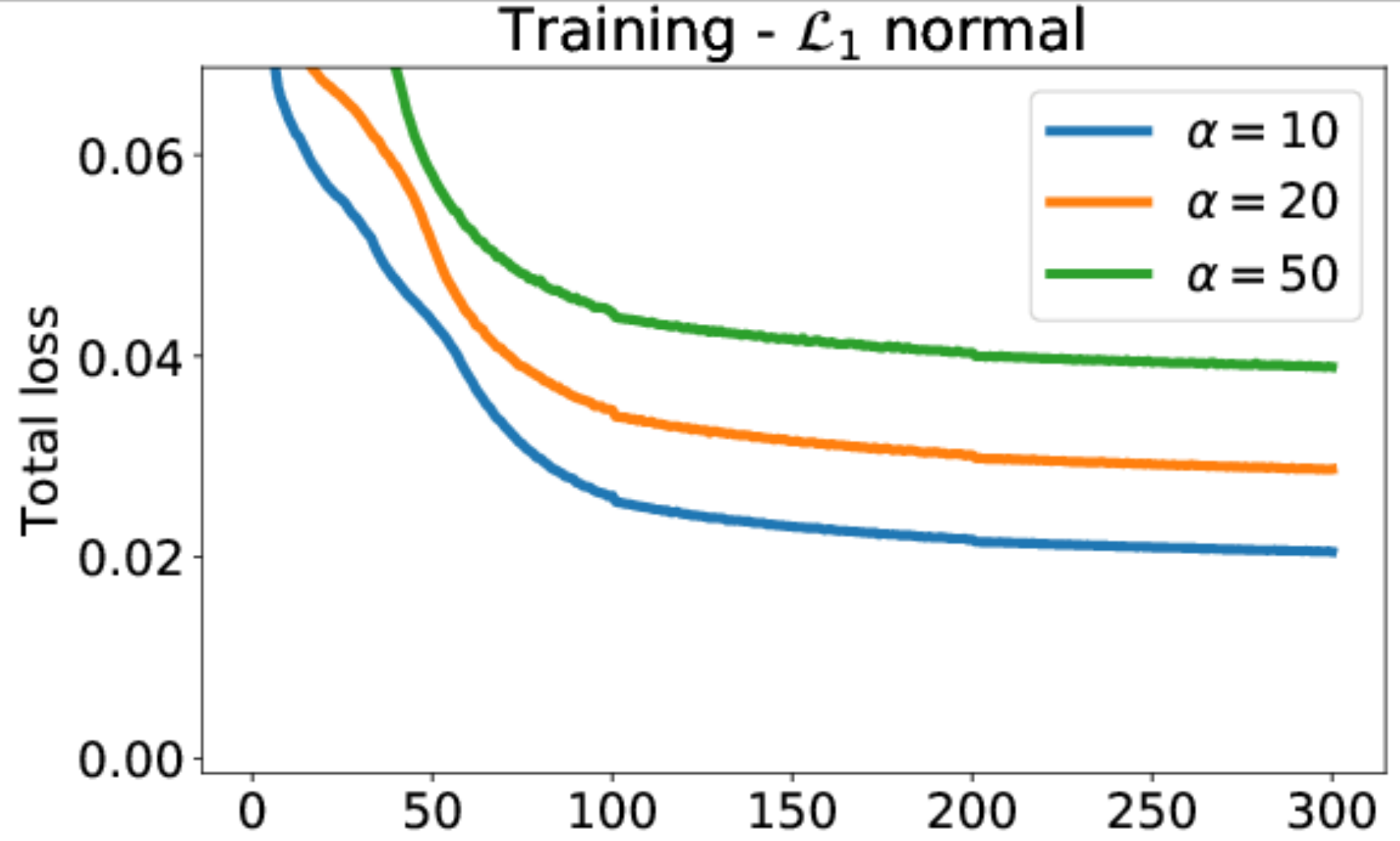}
    \includegraphics[width=0.4\linewidth]{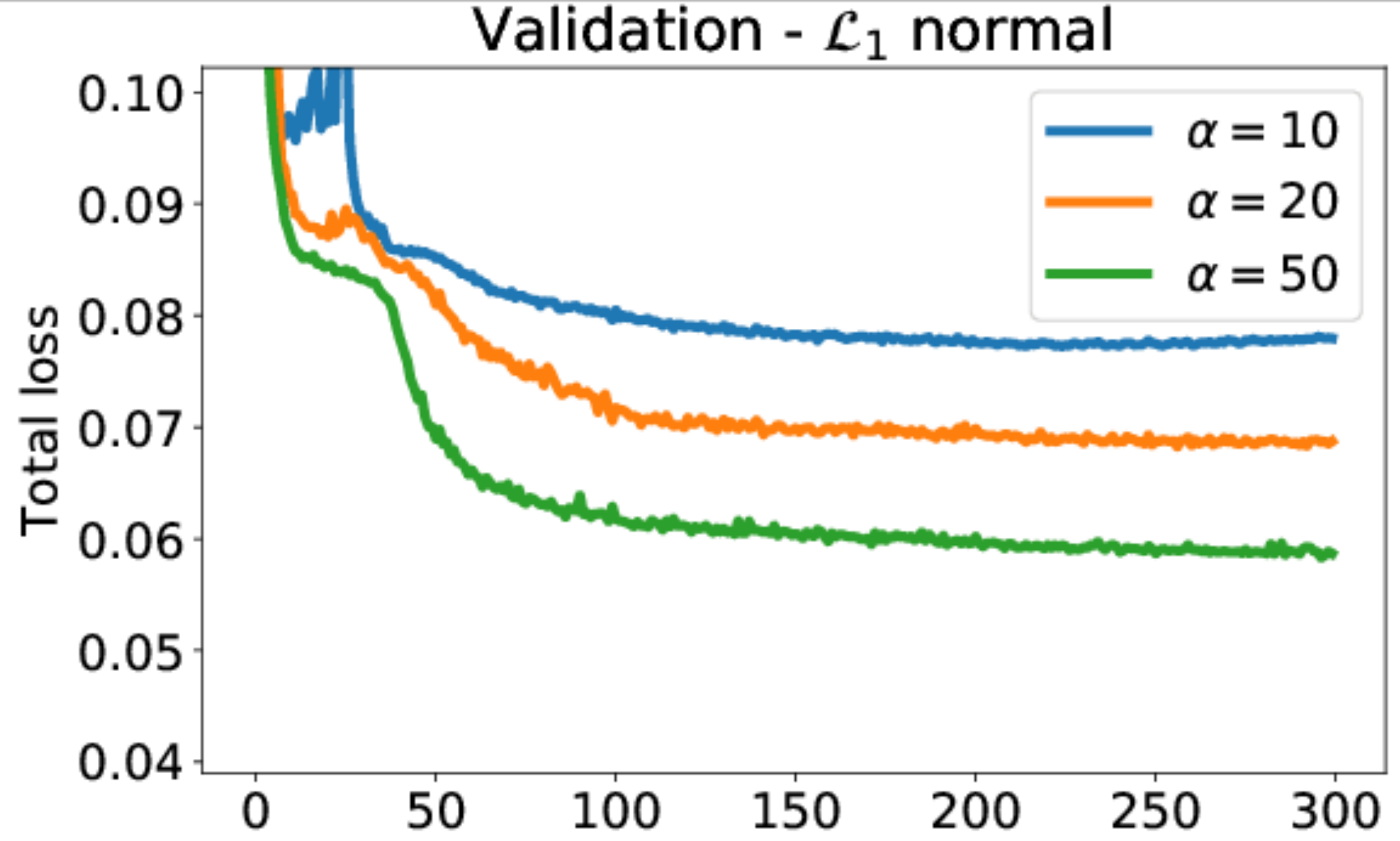}\\
    \includegraphics[width=0.4\linewidth]{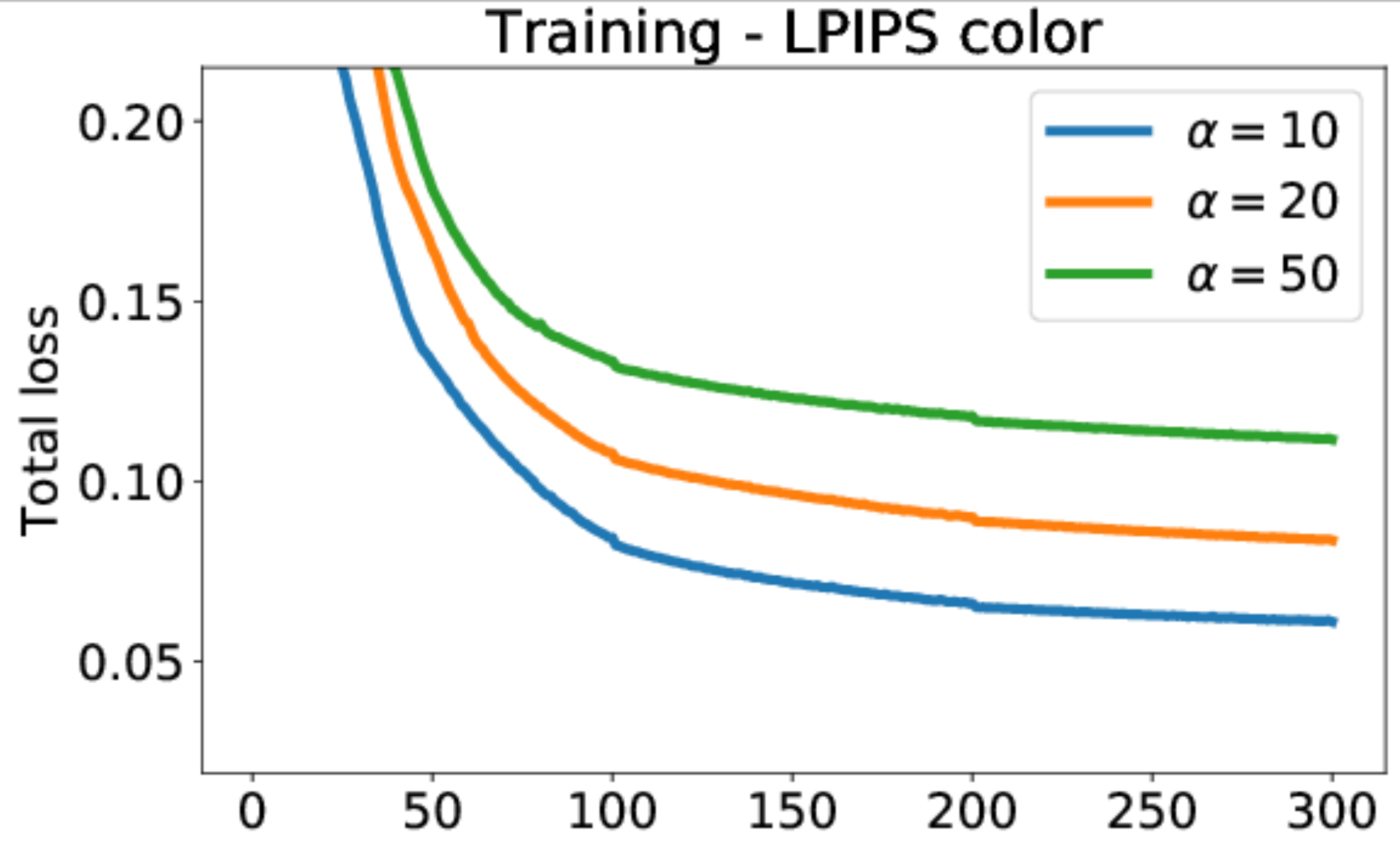}
    \includegraphics[width=0.4\linewidth]{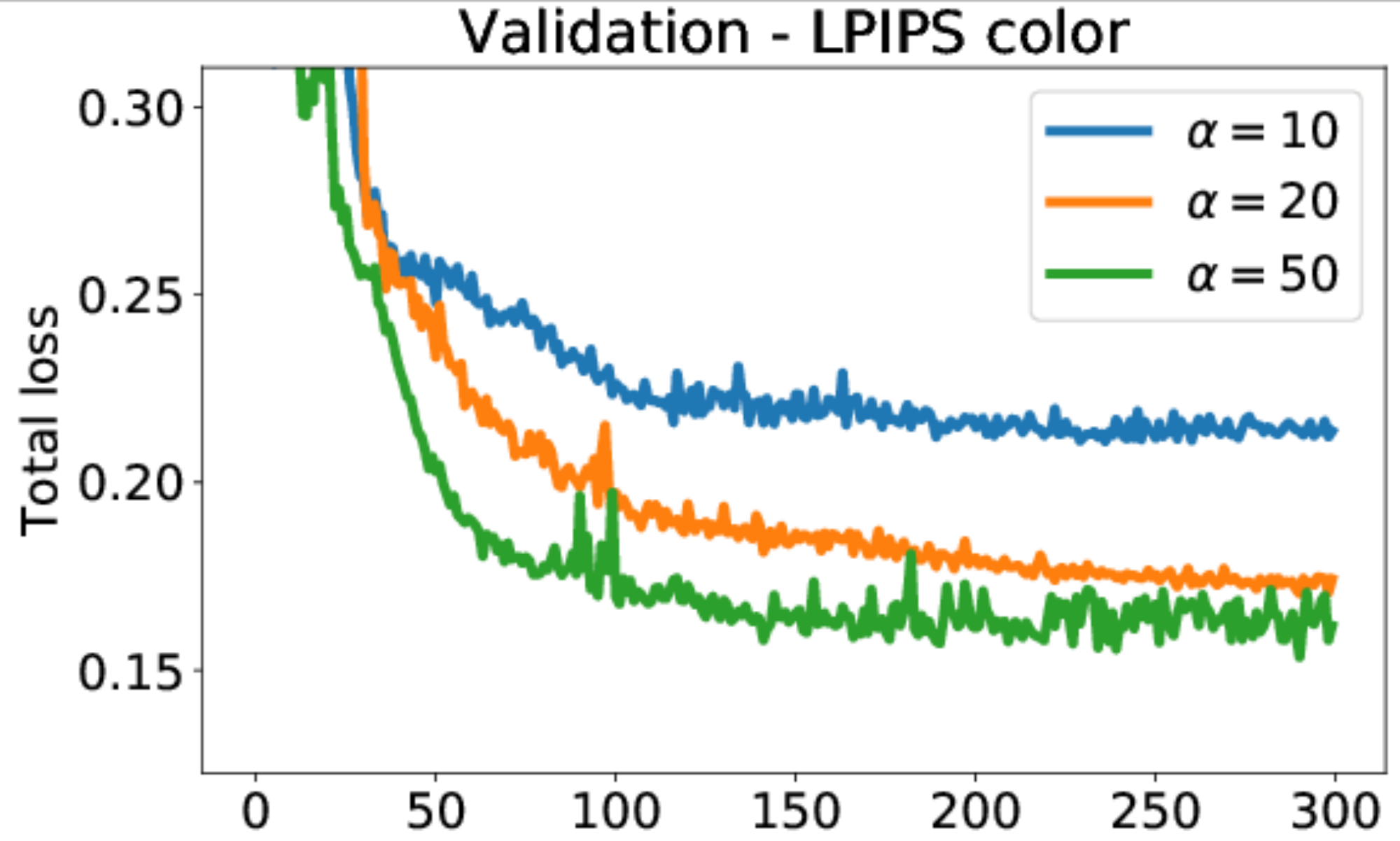}
    \caption{Influence of the sharpness parameter $\alpha$ on the training process. A lower value leads to a lower cost during training, but increases the cost in the validation phase where a perfect step function is used.}
    \label{fig:results:alphaSharpness}
\end{figure}

{\bf Steepness of the Sampling Function}
The parameter $\alpha$ in \autoref{eq:samplingSigmoid} determines the steepness of the sampling function. A perfect step function, as used for testing, is obtained for $\alpha\rightarrow\infty$. \autoref{fig:results:alphaSharpness} compares the total loss on the training and validation data over the course of the optimization for different values of $\alpha$ . A lower value of $\alpha$ leads to a lower cost on the training data, because smoother variations in the fractional samples can be used for reconstruction. However, this behaviour is reversed during validation, because the perfect step function corresponds to a lesser and lesser extent with an increasingly smooth sampling function. Higher values of $\alpha$, on the other hand, lead to better generalization, yet beyond $100$ we observed instabilities in the training as well as numerical precision issues. Therefore, we decided to use $\alpha=50$ in all of our experiments.

\begin{figure}[h]%
    \centering
    \includegraphics[width=\linewidth]{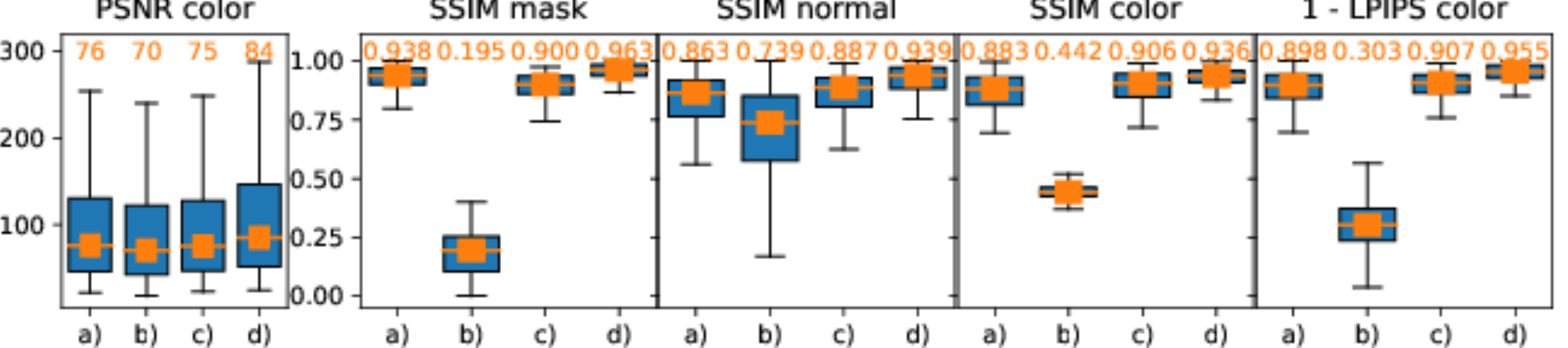}\\%
    \begin{overpic}[width=0.19\linewidth]{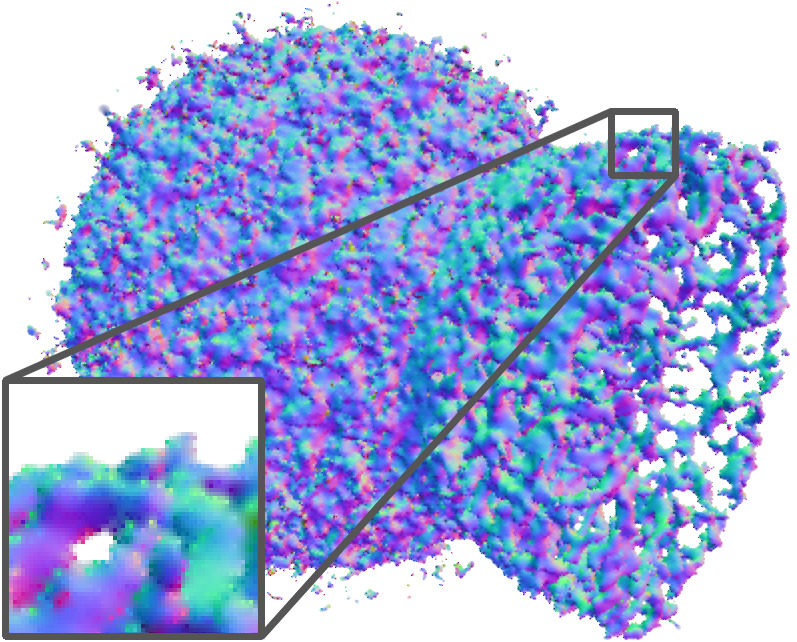}\put(0,70){a)}\end{overpic}~%
    \begin{overpic}[width=0.19\linewidth]{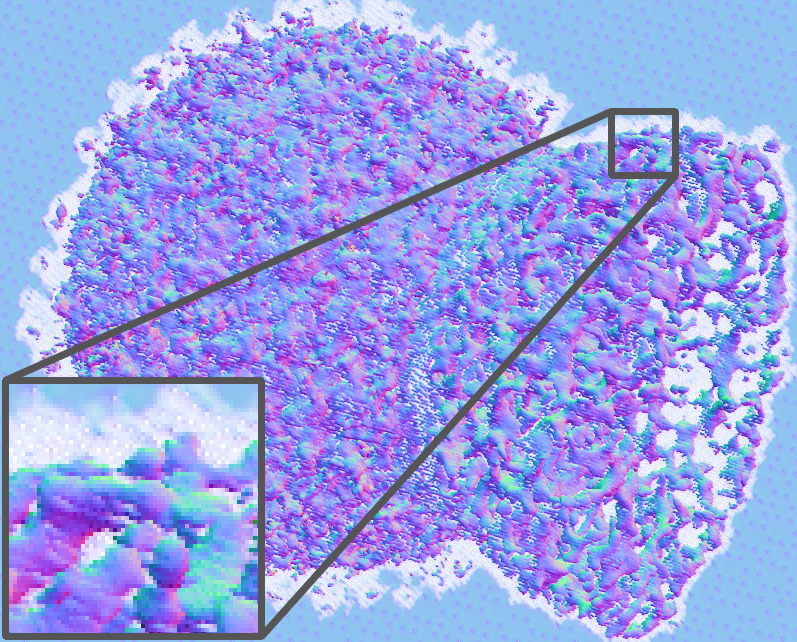}\put(0,70){b)}\end{overpic}~%
    \begin{overpic}[width=0.19\linewidth]{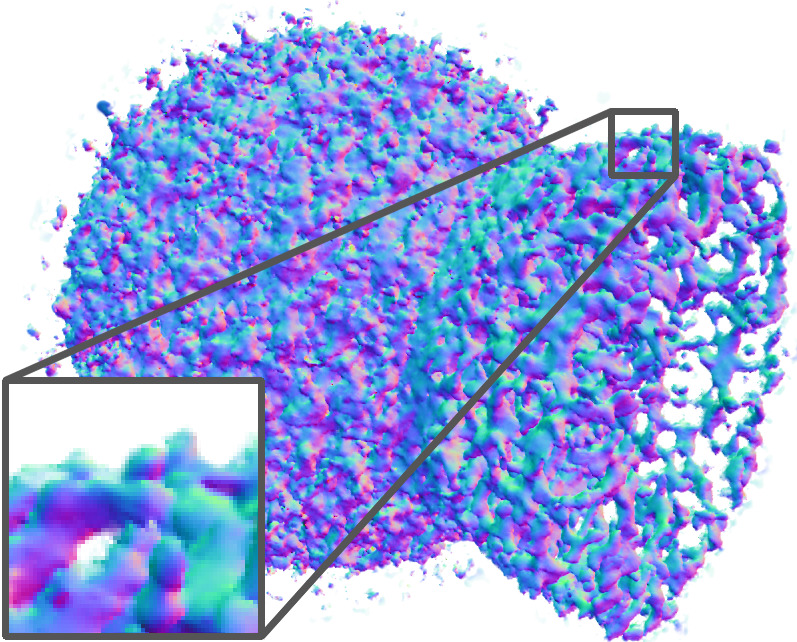}\put(0,70){c)}\end{overpic}~%
    \begin{overpic}[width=0.19\linewidth]{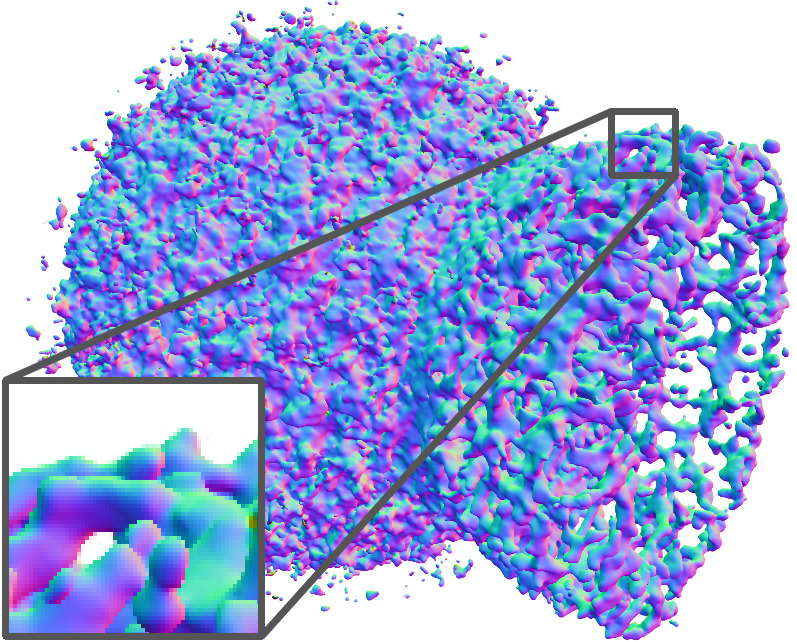}\put(0,70){d)}\end{overpic}%
    \begin{overpic}[width=0.19\linewidth]{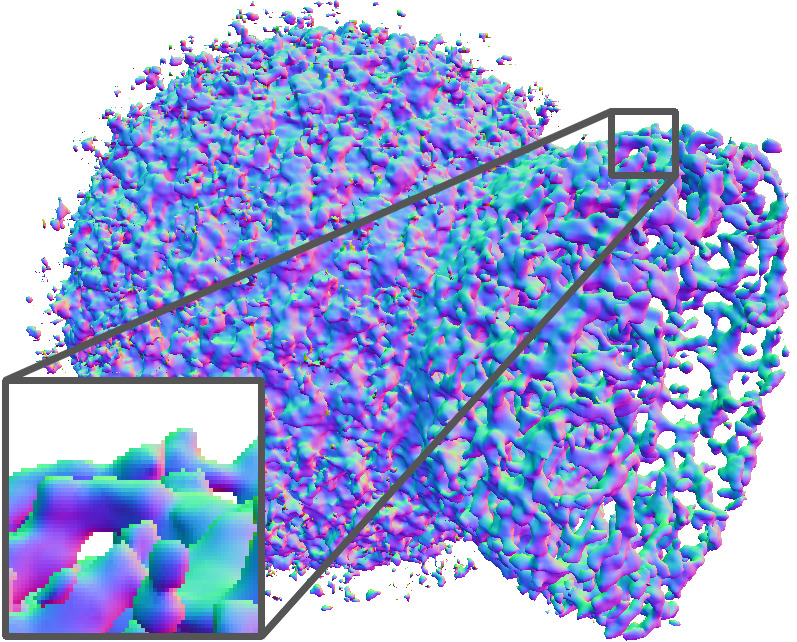}\put(0,70){e)}\end{overpic}%
    \caption{Comparison of different reconstruction methods: (a) Only pull-push-based inpainting. (b) Only network-based reconstruction without residual. (c) Pull-push plus reconstruction network w/o residual. (d) Pull-push inpainting plus reconstruction network with residual. (e) Ground truth. An importance map from screen space gradient magnitudes is used in all examples, with $\mu=5\%$ of samples.}
    \label{fig:results:reconstruction}
\end{figure}

{\bf Residual Connections for Reconstruction}
In principle, there are different options to reconstruct a dense image from the sparse set of samples, including sole inpainting via the pull-push algorithm as well as inpainting in combination with network-based reconstruction w/ or w/o residual connections. In \autoref{fig:results:reconstruction}, the reconstruction quality of all options is compared, using screen space gradient magnitudes as measure for generating the importance map.%

As can be seen, the pull-push algorithm already provides a good initial guess on the reconstructed image, and reconstruction quality reduces significantly when it is not used. On the other hand, the network-based approach fails to reliably fill the empty pixels, which is probably due to the vastly different distances between the sparse samples. When using the pull-push algorithm in combination with network-based reconstruction, but with disabled residual connections, no benefit over sole pull-push-based inpainting is gained. The best result is achieved with both pull-push-based inpainting and residual network connections. This is in line with the findings of Kim~\etAl~\citebody{Kim_2016_CVPR}, that the quality of network-based reconstruction improves if the network needs to learn only the changes to the baseline method. 

\begin{figure}[h]
    \centering
    \includegraphics[width=\linewidth]{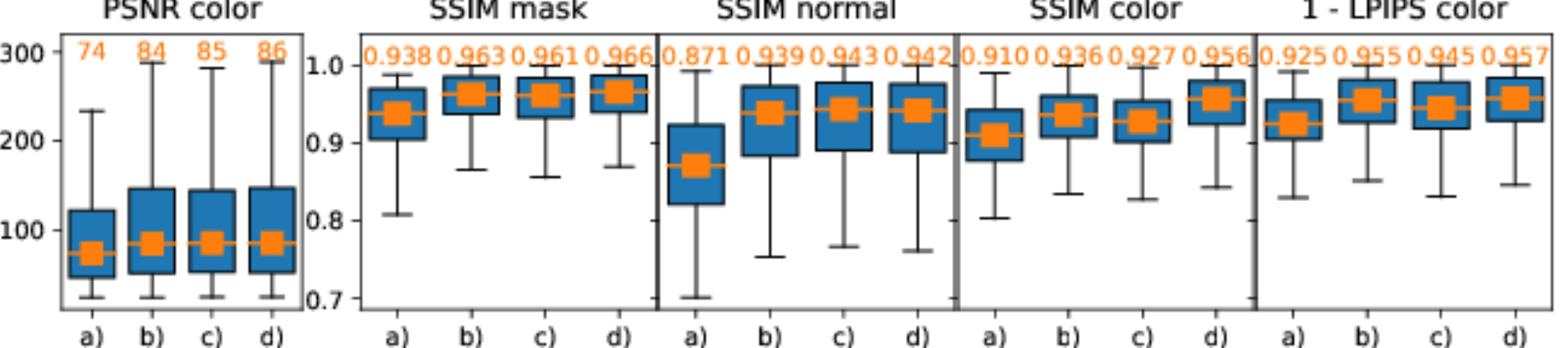}\\%
    \begin{overpic}[width=0.19\linewidth]{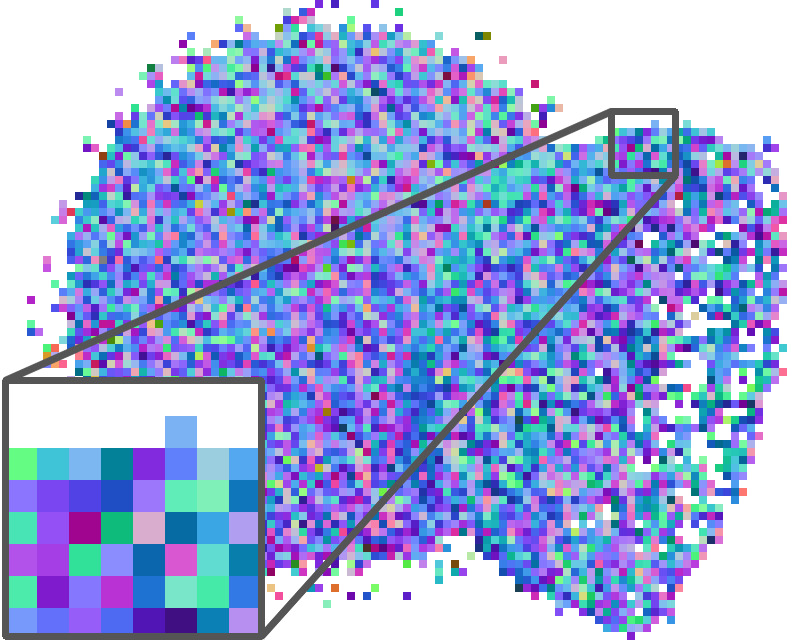}\put(0,70){}\end{overpic}~%
    \begin{overpic}[width=0.19\linewidth]{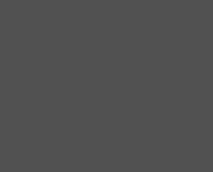}\put(2,65){\color{white}a)}\end{overpic}~%
    \begin{overpic}[width=0.19\linewidth]{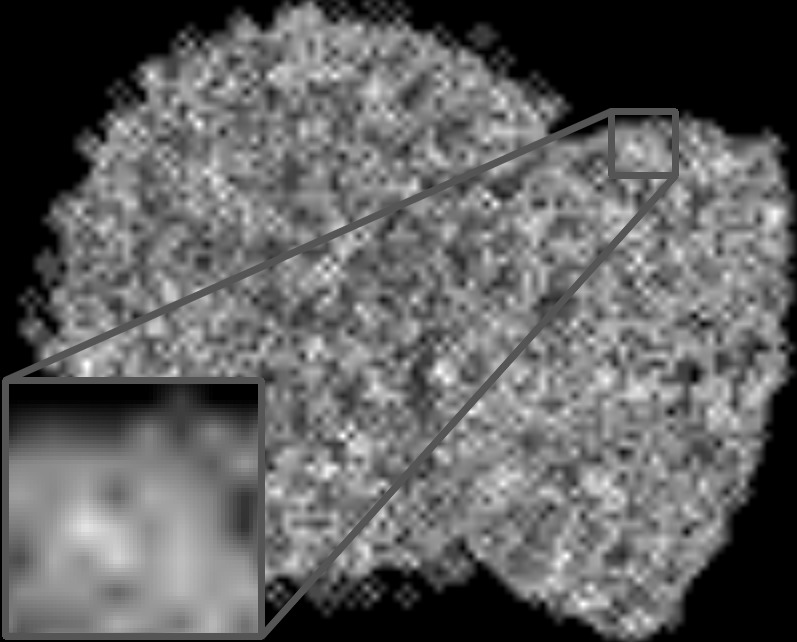}\put(2,65){\color{white}b)}\end{overpic}~%
    \begin{overpic}[width=0.19\linewidth]{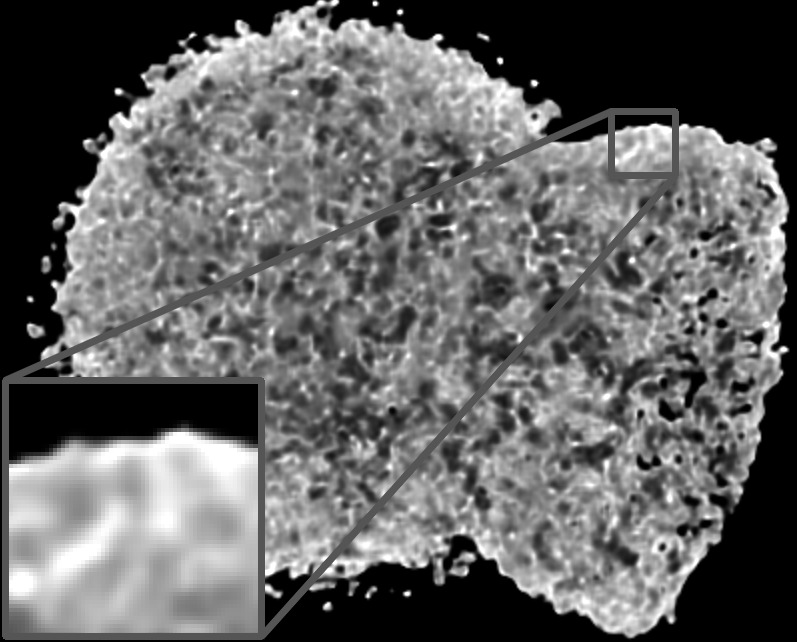}\put(2,65){\color{white}c)}\end{overpic}~%
    \begin{overpic}[width=0.19\linewidth]{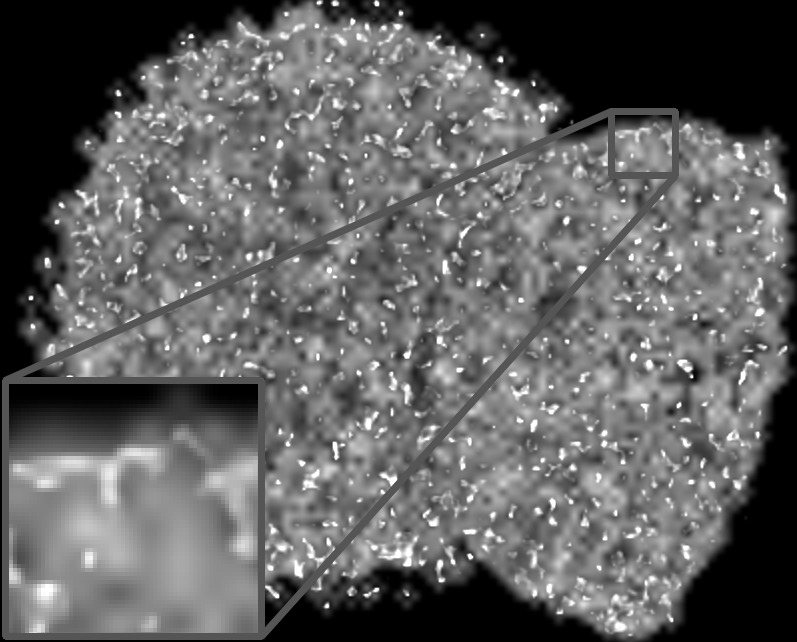}\put(2,65){\color{white}d)}\end{overpic}%
    \caption{Reconstruction quality using different importance maps. Bottom left: Low-resolution input. (a) Constant map, (b) based on gradient magnitudes, (c) only network-based learning, (d) network-based learning with residual on gradient magnitudes. $\mu=5\%$ of samples were used. Top: Quality metrics for options (a) to (d).}
    \label{fig:results:importance}
\end{figure}

{\bf Residual Connections for Importance Mapping}
On the validation data, we then analyze the reconstruction quality using different approaches for generating the importance map, i.e., constant importance, importance derived from screen space gradient magnitudes, as well as network-based importance with or without learning a residual to screen space gradient magnitudes.
\autoref{fig:results:importance} shows the results using the quality metrics described above. 

As expected, screen space gradient magnitudes already hint to some important regions that should be sampled with higher density, significantly outperforming a constant importance map. For reconstructing the mask and normal channels, gradient magnitudes and network-based importance learning differ only marginally w.r.t. reconstruction quality. 
The importance network puts more emphasis on the object silhouettes and leads to an improved reconstruction of the normals over gradient magnitudes.
On the other hand, it is important to note that the network learns the importance of features for an accurate screen space reconstruction without any prior information (\autoref{fig:results:importance}c).
The best results are achieved by combining network-based importance learning and screen space gradient magnitudes via a residual network connection, demonstrating the feasibility of learning features that are important for an accurate reconstruction.

\subsection{Convergence and Regular Sampling}\label{sec:results:sr}
\begin{figure}[t]
    \centering
    \includegraphics[width=\linewidth]{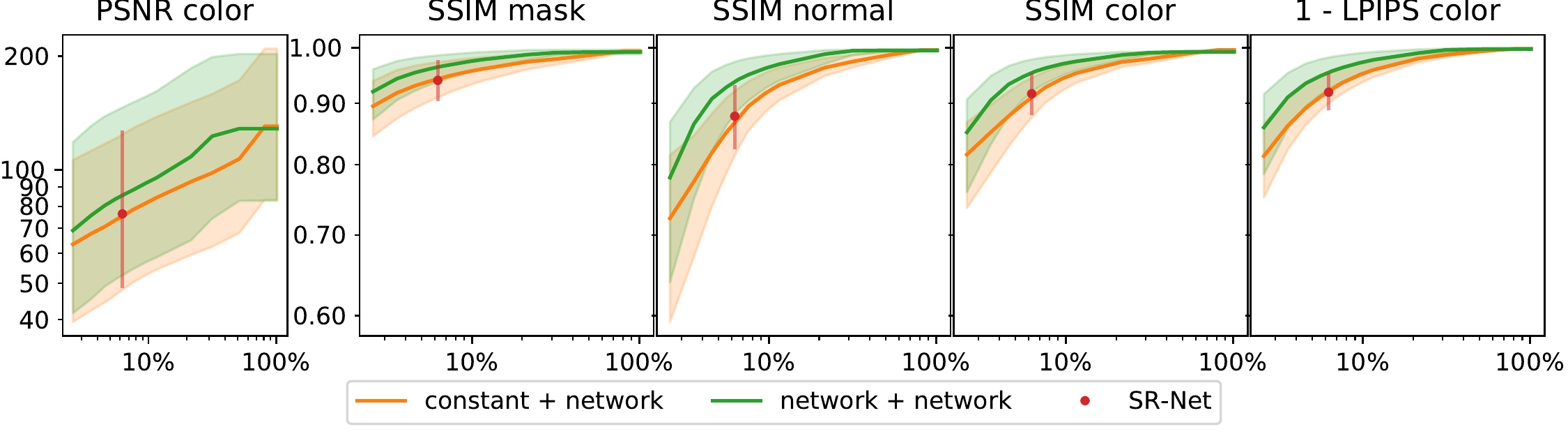}
    \caption[Median reconstruction quality]{Median reconstruction quality and 25\% / 75\% quantiles shown as confidence bands for increasing number of samples. Orange: Network-based pipeline using a constant importance map. Green: Network-based pipeline with the network-based importance map. The red dot represents the 4x-upsampling network from Weiss~\etAl~\citebody{weiss2019isosuperres}.}
    \label{fig:results:heatmap}
\end{figure}

We further analyse the convergence of the proposed sampling pipeline with increasing number of samples. The network is trained with $10\%$ of the samples, but during inference the available number of samples is varied. The results in \autoref{fig:results:heatmap} indicate, that with increasing number of samples the SSIM and LPIPS scores converge against their optima. Even though this seems logical at first, since the reconstruction network modifies the given samples, it could, in principal, converge against some other solution. Notably, already after taking $20\%$ to $30\%$ of samples the reconstruction is very close to the target.

We also compare the quality of adaptive sampling to fixed regular sampling using a 4x-upsampling network ~\citebody{weiss2019isosuperres}. The 4x-upsampling network uses a regular sampling structure comprised of $1/4^2=6.25\%$ of the pixels in the high-resolution image, corresponding to a constant importance map with $6.25\%$ of the samples when adaptive sampling is used. \autoref{fig:results:heatmap} shows that the 4x-upsampling network (red) performs equally good as the adaptive pipeling using a constant importance map (orange). However, when the samples are placed adaptively according to the inferred importance map (green), the reconstruction quality is significantly increased at the same number of samples.

\subsection{Generalizability}\label{sec:results:otherdata}

\begin{figure}[h]
    \centering
    \includegraphics[width=\linewidth]{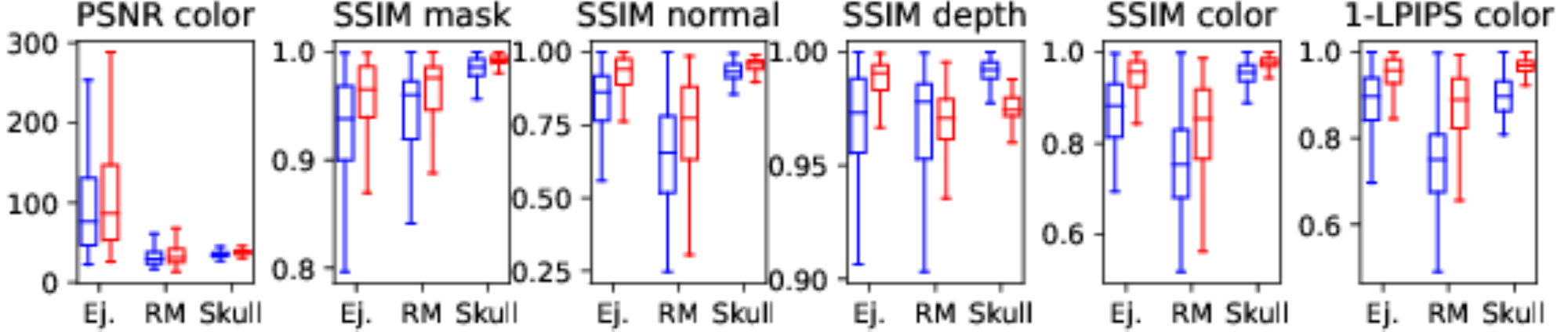}%
    \caption{Quality statistics for novel views of Ejecta and new data sets RM and Skull (see \autoref{fig:result:iso}). Baseline method (blue) refers to gradient magnitude-based importance mapping and pull-push-based inpainting. Results of the proposed network pipeline are shown in red.}
    \label{fig:result:iso-stats}
\end{figure}

The importance and reconstruction networks are trained solely on Ejecta. 
To test how well the networks generalize, they are applied to a number of data sets that were never seen during training. We use a Richtmyer-Meshkov (RM) simulation at $1024\times1024\times960$, CT scans of a human skull (Skull) at $256^3$, an aneurism (Aneurism) at $256^3$, a bug (Bug) at $416^2 \times 247$, and a human body (Thorax) at $256^2 \times 942$, as well as a jet stream simulation (Jet) at $256^3$.
Quantitative statistics for RM, Skull and novel views of Ejecta are given in \autoref{fig:result:iso-stats}. Reconstructed images as well as SSIM and LPIPS statistics for all data sets are shown in \autoref{fig:result:iso}.

The pipeline generalizes well to new data sets and views, and it performs better than the baseline method using gradient magnitude-based importance mapping and pull-push-based inpainting. In particular, the network pipeline produces a tighter spread of the quantitative measures in general, indicating less significant outliers in the reconstructed values.   
The network shows lower scores only for the depth maps reconstructed from sparse samples of RM and Skull. We attribute this to different zoom levels in the renderings and the training images, yet these inaccuracies do not affect the quality of the reconstructed color images.
For reconstruction, we also analyzed the quality of other inpainting algorithms such as PDE-based methods. Notably, these methods are not differentiable and, thus, cannot be used for end-to-end training in combination with the importance network, \changed{yet they can be used for sole sparse image reconstruction. A comparison to the pull-push algorithm, however, does not show any perceivable differences.}
The results further indicate that the network pipeline can reconstruct images at high fidelity from only $5\%$ of the samples that are used to render the data sets at full pixel resolution. In particular sharp edges are well preserved, since the network has learned to increase the sample density along them.

\section{Application to DVR}\label{sec:dvr}

The proposed network pipeline can be applied to images that are rendered via Direct Volume Rendering (DVR), i.e., 
volume ray-casting using an emission-absorption model along the rays of sight. In contrast to isosurface ray-casting, not only one single ray-surface intersection point is rendered, but the colors of many sample points along the rays are blended using $\alpha$-compositing to account for volumetric attenuation.  

\subsection{Training and Validation}
The importance and reconstruction networks receive RGB$\alpha$ images as input, and the network pipeline outputs the reconstructed high-resolution RGB$\alpha$ images. 
Interestingly, we observed a noticeable increase in quality when the gradients at the sample points along the view rays are used by the importance and reconstruction network. 
The normalized gradients in $[-1,1]^3$ along a single ray are treated as emission and blended according to the volume rendering integral, just as blending the RGB colors. The resulting gradient map is then used as an additional input channel.
Since the average gradients indicate, to a certain extent, whether two rays step through vastly different or similar regions, the gradient map serves as an additional coherence indicator. When only a single isosurface is rendered, the resulting values converge against the values in the normal map. 

For training and validation, random transfer functions (TFs) are generated and used to render Ejecta, with $L_1$ losses on color and alpha in combination with a LPIPS-based perceptual loss (\autoref{app:dvr}).
Since the low-resolution input to the importance network is also generated with a TF, the network can learn to select features specific to that TF, even though this was never seen during training. It is important to note that the reconstruction quality strongly depends on the use of TFs that include a broad range of different colors in the training step. For instance, if the training data only contains desaturated colors, strongly saturated colors during testing cannot be reconstructed.

\subsection{DVR Results}
For novel views of Ejecta and the data sets introduced in \autoref{sec:results:otherdata}, \autoref{fig:result:dvr} shows a qualitative analysis of the results of importance sampling and reconstruction using DVR as well as SSIM and LPIPS statistics. None of these data sets was used in the training and validation phases, and the results have been generated using TFs that were never seen during training. The results indicate that the network pipeline generalizes well to new volumes and TFs, yet the reconstruction quality is affected by the occurring color variations. Especially for Thorax and Aneurism, where the TFs introduce rather small scale color variations in some areas, in these areas the network places the samples rather uniformly and, thus, cannot accurately reconstruct the rendered structured. Overall, it can be seen that the reconstruction problem is significantly more challenging when using DVR samples instead of isosurface samples. When rendering isosurfaces, the shading in the interior of the rendered structures is rather smooth, enabling the network to focus on the silhouettes and internal edges. In DVR, on the other hand, the network needs to learn both the shape and the color texture stemming from the application of a TF.

\section{Performance Analysis}

Even though performance improvements are not our main objective, it is interesting to see whether network-based adaptive sampling and image reconstruction can be faster than full-resolution GPU ray-casting, due to the reduced number of samples that need to be taken. The following performance tests were carried out on a workstation running Windows 10 with an Intel Xeon E5-1630 @3.70GHz CPU, 32GB RAM, and an NVidia Titan RTX. 
All timings are averages over \changed{100} frames with random camera positions,
with the screen resolution set to \changed{$1024^2$}. The ray-caster uses a constant step size of 0.25 voxels and tricubic interpolation.

\begin{figure*}[p]
    \centering
    \includegraphics[width=\textwidth]{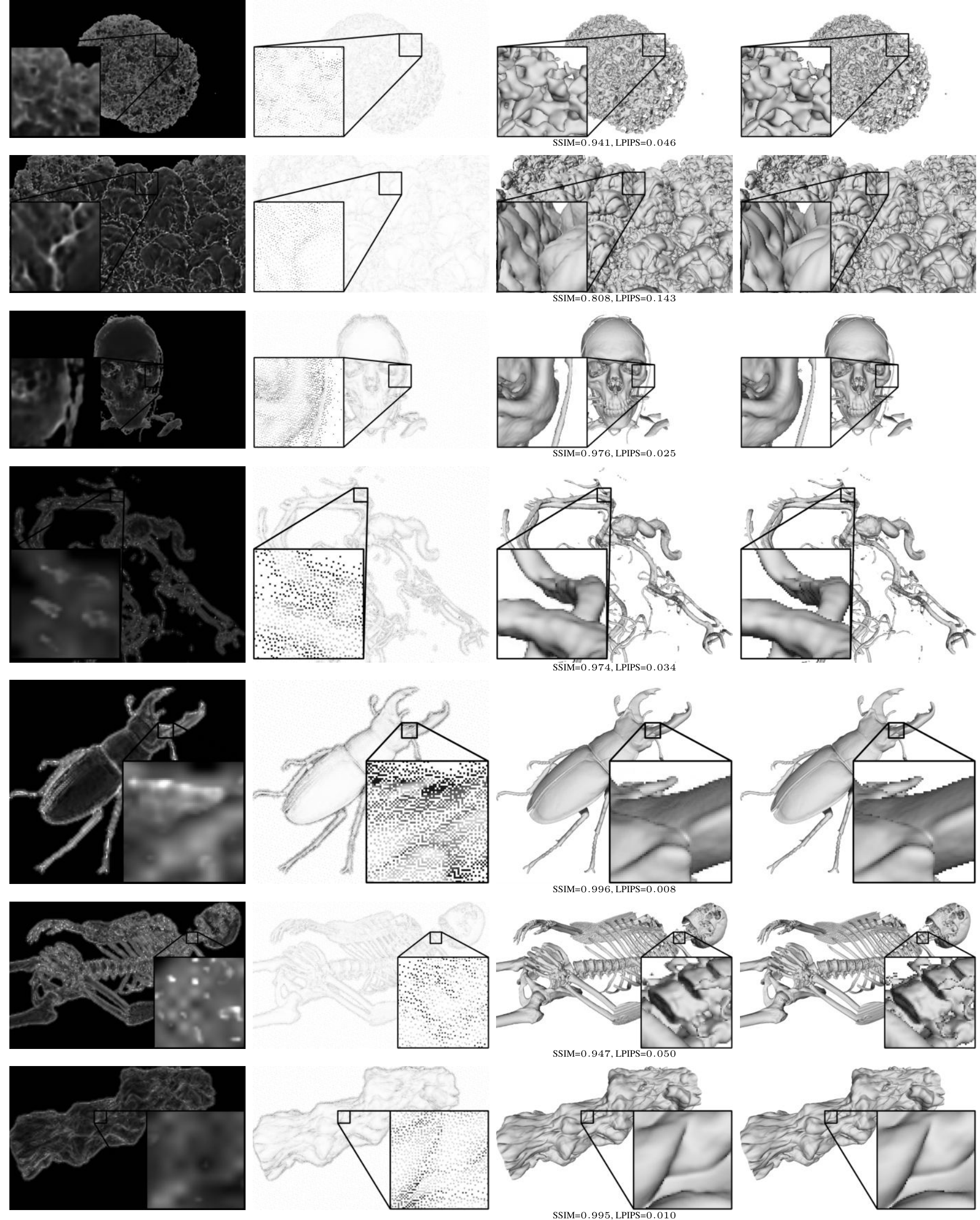}%
    \caption{Visual comparison of adaptive sampling of isosurfaces. From left to right: The importance map and, the sparse set of samples, the inpainted samples, the network output, the ground truth (normals and colors using reconstructed normals). From top to bottom: Novel views of Ejecta, RM, Skull, Aneurism, Bug, Human, Jet. Networks were trained only on Ejecta. Number of samples is $\mu=5\%$ of the pixels in the output image.}
    \label{fig:result:iso}
\end{figure*}

\begin{figure*}[p]
    \centering
    \includegraphics[width=\textwidth]{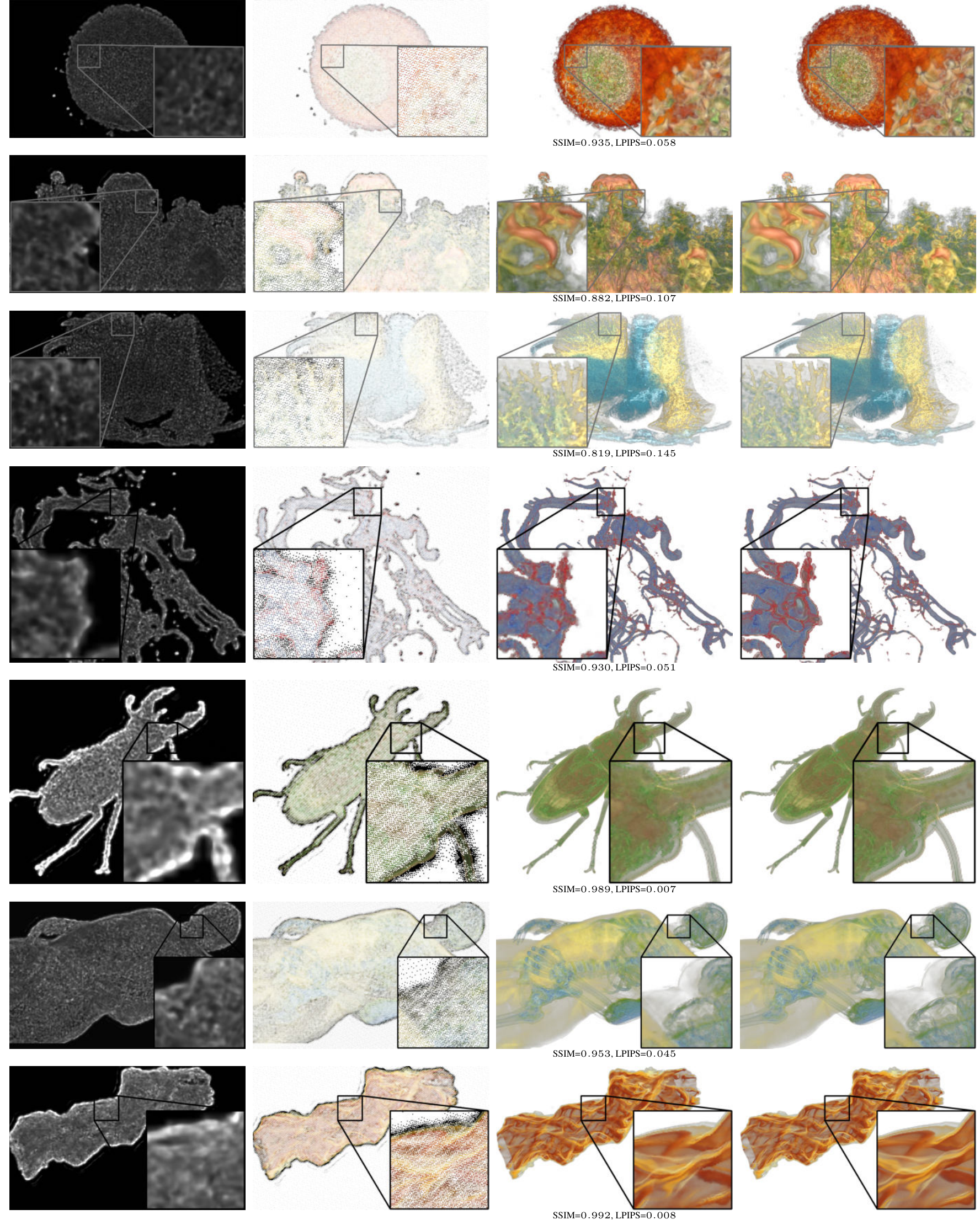}%
    \caption{Visual comparison of adaptive sampling for DVR. Each row shows -- from left to right -- the importance map, the sparse set of samples, the network output, the ground truth. From top to bottom: Novel views of Ejecta, RM, Thorax, Aneurism, Bug, Human, Jet. Networks were trained only on Ejecta. Number of samples is $\mu=10\%$ of the pixels in the output image.}
    \label{fig:result:dvr}
\end{figure*}

For some of the data sets shown in \autoref{fig:result:iso} and \autoref{fig:result:dvr}, \autoref{tab:results:timings} lists the times that are required by the pipeline stages and full-resolution volume ray-casting.
Only for the larger data sets and DVR can the network pipeline achieve a slightly better performance than the ray-caster. Especially the reconstruction network consumes a significant portion of the overall time, sometimes even more than it requires to render at full resolution. This is because the reconstruction network requires a large amount of data access and arithmetic operations on the GPU, independent of the volume resolution. 

\begin{table}[t]
    \caption{Timings (in milliseconds) of network-based volume rendering (averaged over \changed{100} different views at $1024^2$ target resolution) for data sets shown in \autoref{fig:result:iso} and \autoref{fig:result:dvr}. Timings are for rendering the low-resolution input image ($128^2$) and the sparse set of samples ($5\%$ and $10\%$ of the target resolution for isosurface rendering and DVR respectively), generating the importance map and sampling pattern, reconstructing the image, and GPU ray-casting at the target resolution. 
    }
    \label{tab:results:timings}
    \centering
    {\setlength{\tabcolsep}{2pt}\changed{
    \begin{tabular}{cl|cccc}
\multicolumn{2}{c|}{Test case} & Rendering & Importance & Reconstruction & GT \\\hline
\parbox[t]{2mm}{\multirow{3}{*}{\rotatebox[origin=c]{90}{ISO}}}
& RM $1024^3$ & 34.3 & 5.8 & 92.0 & 89.4 \\
& Ejecta $512^3$ & 24.3 & 7.4 & 92.1 & 105.8 \\
& Skull $256^3$ & 6.1 & 5.9 & 93.7 & 27.3 \\\hline
\parbox[t]{2mm}{\multirow{3}{*}{\rotatebox[origin=c]{90}{DVR}}}
& RM $1024^3$ & 51.7 & 5.8 & 91.8 & 158.3 \\
& Ejecta $512^3$ & 46.5 & 5.8 & 92.2 & 224.8 \\
& Thorax $512^3$ & 15.9 & 5.9 & 92.9 & 63.7 \\
\end{tabular}}
    }
\end{table}

\begin{table}[tbh]
    \caption{\changed{Performance scaling w.r.t. image and volume size. Each entry shows the total time of the network pipeline (low-resolution rendering, importance network, sparse sampling, reconstruction network) and the time required by the volume ray-caster at full resolution. All timings are in milliseconds. The cells are colored with a diverging color map, encoding the performance differences from red (superior performance of ray-casting) to blue (superior performance of the network pipeline).}}
    \label{tab:performance:scaling}
    \centering
    \setlength{\tabcolsep}{2pt}
\begin{tabular}{cc|cl|c|c|c|c}
&&&& \multicolumn{4}{c}{Screen resolution} \\
&&&& 256 & 512 & 1024 & 2048 \\ \hline
\parbox[t]{2mm}{\multirow{8}{*}{\rotatebox[origin=c]{90}{ISO}}}%
& \parbox[t]{2mm}{\multirow{4}{*}{\rotatebox[origin=c]{90}{Ejecta}}}%
& \parbox[t]{2mm}{\multirow{16}{*}{\rotatebox[origin=c]{90}{Volume resolution}}}%
&256 & \cellcolor[rgb]{1.00, 0.63, 0.63} $24/11$  & \cellcolor[rgb]{1.00, 0.61, 0.61} $45/20$  & \cellcolor[rgb]{1.00, 0.79, 0.79} $115/75$  & \cellcolor[rgb]{1.00, 0.85, 0.85} $398/294$  \\ \cline{4-8} 
&&&512 & \cellcolor[rgb]{1.00, 0.68, 0.68} $31/16$  & \cellcolor[rgb]{1.00, 0.72, 0.72} $55/31$  & \cellcolor[rgb]{1.00, 0.92, 0.92} $124/106$  & \cellcolor[rgb]{1.00, 1.00, 1.00} $405/400$  \\ \cline{4-8} 
&&&1024 & \cellcolor[rgb]{0.98, 0.98, 1.00} $45/47$  & \cellcolor[rgb]{0.82, 0.82, 1.00} $73/104$  & \cellcolor[rgb]{0.78, 0.78, 1.00} $141/221$  & \cellcolor[rgb]{0.81, 0.81, 1.00} $445/655$  \\ \cline{4-8} 
&&&2048 & \cellcolor[rgb]{0.46, 0.46, 1.00} $85/278$  & \cellcolor[rgb]{0.28, 0.28, 1.00} $132/806$  & \cellcolor[rgb]{0.30, 0.30, 1.00} $238/1358$  & \cellcolor[rgb]{0.47, 0.47, 1.00} $724/2324$  \\ \cline{4-8} 
\cline{2-2}
 
& \parbox[t]{2mm}{\multirow{4}{*}{\rotatebox[origin=c]{90}{RM}}}%
&&256 & \cellcolor[rgb]{1.00, 0.28, 0.28} $24/4$  & \cellcolor[rgb]{1.00, 0.42, 0.42} $41/11$  & \cellcolor[rgb]{1.00, 0.55, 0.55} $111/42$  & \cellcolor[rgb]{1.00, 0.61, 0.61} $393/173$  \\ \cline{4-8} 
&&&512 & \cellcolor[rgb]{1.00, 0.32, 0.32} $31/6$  & \cellcolor[rgb]{1.00, 0.45, 0.45} $51/15$  & \cellcolor[rgb]{1.00, 0.67, 0.67} $116/58$  & \cellcolor[rgb]{1.00, 0.75, 0.75} $397/236$  \\ \cline{4-8} 
&&&1024 & \cellcolor[rgb]{1.00, 0.49, 0.49} $45/15$  & \cellcolor[rgb]{1.00, 0.60, 0.60} $71/31$  & \cellcolor[rgb]{1.00, 0.81, 0.81} $132/89$  & \cellcolor[rgb]{1.00, 0.94, 0.94} $411/366$  \\ \cline{4-8} 
&&&2048 & \cellcolor[rgb]{0.83, 0.83, 1.00} $87/122$  & \cellcolor[rgb]{0.67, 0.67, 1.00} $124/249$  & \cellcolor[rgb]{0.64, 0.64, 1.00} $211/453$  & \cellcolor[rgb]{0.89, 0.89, 1.00} $646/802$  \\ \Cline{3pt}{4-8} 
\Cline{3pt}{1-2}
 
\parbox[t]{2mm}{\multirow{8}{*}{\rotatebox[origin=c]{90}{DVR}}}%
& \parbox[t]{2mm}{\multirow{4}{*}{\rotatebox[origin=c]{90}{Ejecta}}}%
&&256 & \cellcolor[rgb]{1.00, 0.54, 0.54} $46/17$  & \cellcolor[rgb]{1.00, 0.84, 0.84} $64/46$  & \cellcolor[rgb]{0.89, 0.89, 1.00} $131/162$  & \cellcolor[rgb]{0.84, 0.84, 1.00} $413/574$  \\ \cline{4-8} 
&&&512 & \cellcolor[rgb]{1.00, 0.76, 0.76} $59/36$  & \cellcolor[rgb]{1.00, 0.99, 0.99} $75/73$  & \cellcolor[rgb]{0.78, 0.78, 1.00} $144/225$  & \cellcolor[rgb]{0.72, 0.72, 1.00} $447/785$  \\ \cline{4-8} 
&&&1024 & \cellcolor[rgb]{0.72, 0.72, 1.00} $86/151$  & \cellcolor[rgb]{0.56, 0.56, 1.00} $103/262$  & \cellcolor[rgb]{0.59, 0.59, 1.00} $180/433$  & \cellcolor[rgb]{0.64, 0.64, 1.00} $594/1274$  \\ \cline{4-8} 
&&&2048 & \cellcolor[rgb]{0.43, 0.43, 1.00} $160/575$  & \cellcolor[rgb]{0.20, 0.20, 1.00} $186/1692$  & \cellcolor[rgb]{0.21, 0.21, 1.00} $333/2771$  & \cellcolor[rgb]{0.47, 0.47, 1.00} $1319/4292$  \\ \cline{4-8} 
\cline{2-2}
 
& \parbox[t]{2mm}{\multirow{4}{*}{\rotatebox[origin=c]{90}{RM}}}%
&&256 & \cellcolor[rgb]{1.00, 0.40, 0.40} $32/8$  & \cellcolor[rgb]{1.00, 0.60, 0.60} $49/21$  & \cellcolor[rgb]{1.00, 0.75, 0.75} $117/71$  & \cellcolor[rgb]{1.00, 0.82, 0.82} $398/280$  \\ \cline{4-8} 
&&&512 & \cellcolor[rgb]{1.00, 0.42, 0.42} $41/11$  & \cellcolor[rgb]{1.00, 0.64, 0.64} $58/27$  & \cellcolor[rgb]{1.00, 0.87, 0.87} $125/96$  & \cellcolor[rgb]{1.00, 0.95, 0.95} $404/363$  \\ \cline{4-8} 
&&&1024 & \cellcolor[rgb]{1.00, 0.65, 0.65} $60/29$  & \cellcolor[rgb]{1.00, 0.83, 0.83} $81/58$  & \cellcolor[rgb]{0.97, 0.97, 1.00} $149/158$  & \cellcolor[rgb]{0.90, 0.90, 1.00} $450/549$  \\ \cline{4-8} 
&&&2048 & \cellcolor[rgb]{0.72, 0.72, 1.00} $115/203$  & \cellcolor[rgb]{0.52, 0.52, 1.00} $143/410$  & \cellcolor[rgb]{0.55, 0.55, 1.00} $248/656$  & \cellcolor[rgb]{0.78, 0.78, 1.00} $830/1304$  
 
\end{tabular}
\end{table}

\changed{On the one hand, the performance of the reconstruction network scales linearly with the number of pixels, and hence quadratically with the screen resolution. Volume rendering, on the other hand, scales quadratically with the screen resolution but also linear in the volume resolution. The sampling stage, even though it also scales in the volume resolution, performs a significantly smaller number of sampling operations than the full-resolution ray-caster. Thus, its overall contribution is negligible, so that performance benefits can be expected with increasing image and volume size.
This is demonstrated in \autoref{tab:performance:scaling}, where versions of RM and Ejecta at $2048^3$ are rendered at different resolution levels and large images sizes. Note that in these experiments an Nvidia Titan RTX graphics card with 24GB of memory was used to keep all data in memory.
}

\changed{It can be seen that for large image sizes---where the GPU is fully utilized by the network---and volume sizes larger than $1024^3$, the network pipeline outperforms the GPU ray-caster. Even though a ray-caster using advanced acceleration schemes can achieve improved performance, we are confident that in these scenarios faster deep-learning hardware and performance-optimized network architectures will let the performance differences grow due to better scalability of the network pipeline. 
}



\section{Conclusion and Future Work}\label{sec:conclusion}

In this paper we have introduced and analyzed a network pipeline that learns adaptive screen space sampling and reconstruction for 3D visualization, with the focus on volume rendering applications. For the first time, to our best knowledge, a fully differentiable adaptive sampling pipeline comprised of an importance network, a sampling stage, and a reconstruction network is proposed. Our experiments have shown, that the pipeline learns to determine the locations that are important for an accurate image reconstruction, and achieves high reconstruction quality for a sparse set of samples. 

We are particular intrigued about the quality of the results compared to sampling methods that consider explicitly certain feature descriptors. Even without such supervision, the network pipeline can improve on the reconstruction quality, using solely image-based quality losses. We believe
that especially for data visualization there is value in the observation that artificial neural networks can learn the relevance of structures for generating visual representations. 
For sole rendering tasks, \changed{on the other hand, superior performance compared to classical volume ray-casting can only be achieved for large image and volume sizes.}

\changed{The application to DVR opens the interesting question whether the proposed network pipeline can be used beyond adaptive sampling in screen space, and learn where to sample in object space so that the relevant information is conveyed visually. Conceptually this requires end-to-end learning of a mapping from a low-resolution object space representation to a high-resolution visual representation. The ultimate goal is to let the network learn to convert a low-resolution input volume to a compact yet feature-preserving latent-space representation from which a highly accurate view can be inferred.}

In particular, we envision a neural volume rendering pipeline, where during training a neural scene representation is build and trained end-to-end with a renderer that learns sampling and color mapping simultaneously. In the future, we will analyse whether a network can learn a suitable color mapping for a given volumetric field. We also see challenging research problems in the area of transfer learning, to infer the most important samples for training, and to generate synthetic volumetric fields to enable training in domains where training data is rare.


\bibliographystylebody{abbrv-doi-narrow}
\bibliographybody{main}

\clearpage
\appendices

\section{Comparison with the U-Net for Reconstruction}\label{app:unet}
For reconstruction, we also tested different variants (by varying the number of levels and channels at each level) of the U-Net architecture~\citeapp{ronneberger2015UNet-App}. As one can see in \autoref{fig:results:unet}, in our application the EnhanceNet vastly outperforms all considered U-Net variants.

\begin{figure}[h]
    \centering
    \resizebox{0.6\linewidth}{!}{\begin{tikzpicture}[scale=0.6,
	every text node part/.style={align=center},
	every node/.style={minimum size=0.7cm,inner sep=1pt},
	every label/.style={shape=rectangle, draw=none, fill=none, minimum size=0pt,inner sep=2pt},
	data/.style={draw,rectangle,fill=blue!20},
	op/.style={draw,rectangle,inner sep=2pt,fill=yellow!30,rotate=90,minimum size=0cm,minimum width=2.7cm},
	base/.style={draw,rectangle,inner sep=2pt,fill=yellow!30,rotate=90,minimum size=0cm},
	netA/.style={draw,rectangle,inner sep=2pt,fill=red!40,rotate=90,minimum size=0cm,minimum width=2.8cm},
	netB/.style={draw,rectangle,inner sep=2pt,fill=red!40,rotate=90,minimum size=0.2cm,minimum width=1.4cm},
	netC/.style={draw,rectangle,inner sep=2pt,fill=red!40,rotate=90,minimum size=0.3cm,minimum width=0.7cm},
	netD/.style={draw,rectangle,inner sep=2pt,fill=red!40,rotate=90,minimum size=0.5cm,minimum width=0.35cm},
	netE/.style={draw,rectangle,inner sep=2pt,fill=red!40,rotate=90,minimum size=0.9cm,minimum width=0.175cm},
	dataflow/.style={very thick,->},
	tmpflow/.style={very thick,->,yellow!40!black,dashed}]
	
\node[data] (I1) at (-3,0.5) {$S$};

\node[op,minimum width=0cm] (Imp) at (-1,0) {Inpainting};

\node[draw,circle,fill=none,minimum size=0.2cm] (S1) at (0,0.5) {};
\node[draw,circle,fill=none,minimum size=0.2cm] (S3) at (0.8,1) {};
\node[draw,circle,fill=none,minimum size=0.2cm] (S2) at (0,1.5) {};

\node[netA] (C1a) at (2,0) {};
\node[netA] (C1b) at (2.4,0) {};
\node[black!70] at (1.5,-2) {64};

\node[netB] (C2a) at (2.4,-4) {};
\node[netB] (C2b) at (2.9,-4) {};
\node[black!70] at (1.7,-4.9) {128};

\node[netC] (C3a) at (2.9,-6.1) {};
\node[netC] (C3b) at (3.6,-6.1) {};
\node[black!70] at (2.1,-6.5) {256};

\node[netD] (C4a) at (3.6,-7.3) {};
\node[netD] (C4b) at (4.7,-7.3) {};
\node[black!70] at (2.6,-7.4) {512};

\node[netE] (C5a) at (4.7,-8.2) {};
\node[netE] (C5b) at (7.5,-8.2) {};
\node[black!70] at (3.3,-8.2) {1024};

\node[netD] (C4c) at (7.5,-7.3) {};
\node[netD] (C4d) at (8.6,-7.3) {};

\node[netC] (C3c) at (8.6,-6.1) {};
\node[netC] (C3d) at (9.3,-6.1) {};

\node[netB] (C2c) at (9.3,-4) {};
\node[netB] (C2d) at (9.8,-4) {};

\node[netA] (C1c) at (9.8,0) {};
\node[netA] (C7) at (10.5,0) {Conv, $C$};

\node[draw,circle,fill=none,minimum size=0.2cm] (S4) at (9.5,3) {};
\node[draw,circle,fill=none,minimum size=0.2cm] (S5) at (10.5,3) {};
\node[minimum size=0,circle,draw,inner sep=0pt] (A2) at (11.7,0) {\LARGE $+$};

\node[data] (Out) at (13.5,0) {$O$};

\draw[dataflow] (I1) -- (-2,0.5) -- (-2,0) -- (Imp);
\draw[dataflow] (Imp) -- (-0.5,0) -- (-0.5,0.5) -- (S1);
\draw[dataflow] (I1) -- (-2,0.5) -- (-2,2) -- (-0.5,2) -- (-0.5,1.5) -- (S2);
\draw[thick] (S2) -- (S3);
\draw[thick,dashed] (S1) -- (S3);
\draw[dataflow] (S3) -- (1.3,1) -- (C1a);

\draw[dataflow] (C1a) -- (C1b);
\draw[dataflow] (C1b) -- (C1c);
\draw[dataflow] (C1c) -- (C7);
\draw[dataflow,red!70!black] (C1b) -- (C2a);
\draw[dataflow] (C2a) -- (C2b);
\draw[dataflow] (C2b) -- (C2c);
\draw[dataflow] (C2c) -- (C2d);
\draw[dataflow,green!70!black] (C2d) -- (C1c);
\draw[dataflow,red!70!black] (C2b) -- (C3a);
\draw[dataflow] (C3a) -- (C3b);
\draw[dataflow] (C3b) -- (C3c);
\draw[dataflow] (C3c) -- (C3d);
\draw[dataflow,green!70!black] (C3d) -- (C2c);
\draw[dataflow,red!70!black] (C3b) -- (C4a);
\draw[dataflow] (C4a) -- (C4b);
\draw[dataflow] (C4b) -- (C4c);
\draw[dataflow] (C4c) -- (C4d);
\draw[dataflow,green!70!black] (C4d) -- (C3c);
\draw[dataflow,red!70!black] (C4b) -- (C5a);
\draw[dataflow] (C5a) -- (C5b);
\draw[dataflow,green!70!black] (C5b) -- (C4c);

\draw[dataflow] (C7) -- (A2);
\draw[dataflow] (S5) -- (11.7,3) -- (A2);
\draw[dataflow] (S3) -- (1.3,1) -- (1.3,3) -- (S4);
\draw[dataflow] (A2) -- (Out);

\draw[thick] (9.7,3.3) -- (S5);
\node[] at (10,3.8) {global residual?};

\node[netC] at (12,-4) {};
\node[black!70] at (11.5,-4.5) {X};
\node at (15.5,-4) {Conv, X channels; ReLU};
\draw[dataflow,red!70!black] (12,-4.9) -- (12,-5.5);
\node at (14,-5.2) {max pool 2x2};
\draw[dataflow,green!70!black] (12,-5.9) -- (12,-6.5);
\node at (14.7,-6.2) {bilinear upsampling};

\end{tikzpicture}}
    \mycaptionspacing%
    \caption{Reconstruction network based on the U-Net architecture. See \autoref{fig:architectures:recEnhance} for the EnhanceNet  architecture we use in our experiments.}
    \label{fig:my_label}
\end{figure}
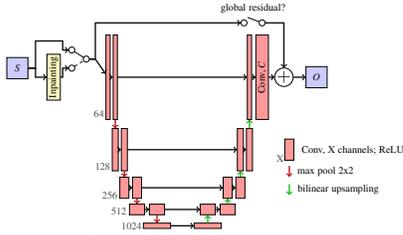
\begin{figure}[h]
    \centering
    \includegraphics[width=\linewidth]{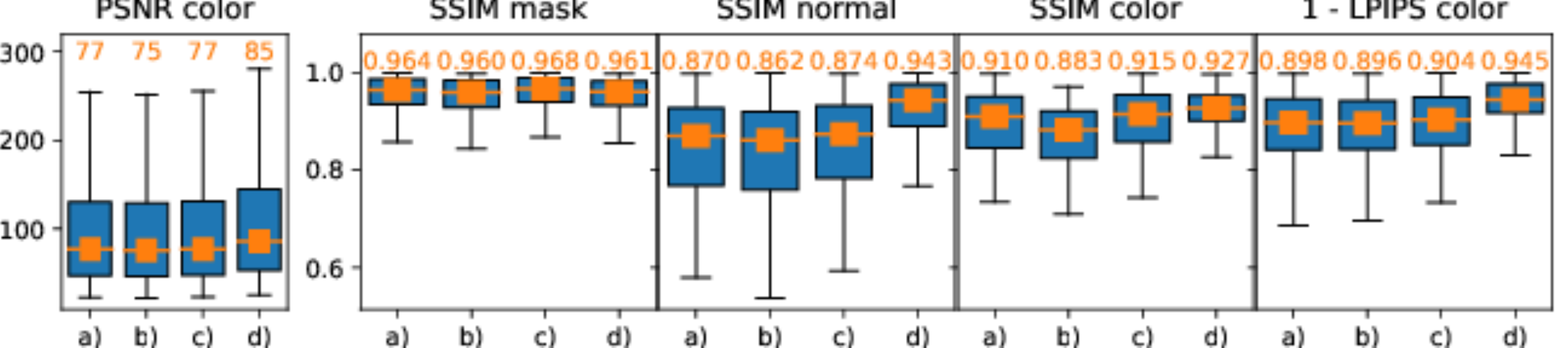}%
    \mycaptionspacing%
    \caption{Comparison of the U-Net and EnhanceNet for sparse image reconstruction: U-Net 4-4 (a), U-Net 5-3 (b), U-Net 5-4 (c), EnhanceNet (d). $a$-$b$ indicates $a$ levels and $2^{b+i}$ channels in level $i$ (zero-based). The importance network is trained together with the reconstruction network.}
    \label{fig:results:unet}
\end{figure}

\section{Comparison of Different Sampling Pattern}\label{app:sampling}

\begin{figure}%
\centering%
\includegraphics[width=\columnwidth]{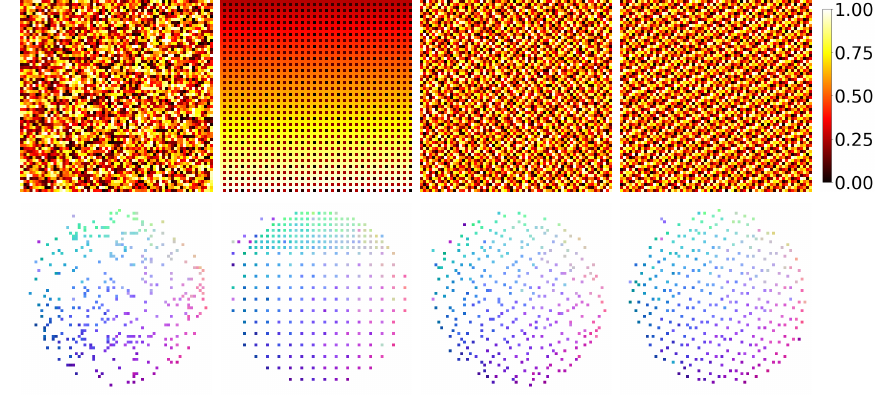}~%
\mycaptionspacing%
\caption{Comparison of random sampling, regular sampling, Halton sampling and plastic sampling (left to right). Top: The sampling sequences. Bottom: The sequences applied to render a sphere with constant importance of $\mu=0.1$ (shown are color coded normals at rendered fragments).}%
\label{fig:samplingComparison}%
\end{figure}

For deterministic and parallelizable sampling on the GPU, we use a pre-computed sampling pattern in combination with rejection sampling (\autoref{sec:method:sampling}).
The sampling pattern $P\in[0,1]^{H\times W}$ contains permutations of uniformly distributed numbers in $[0,1]$, $\frac{1}{HW}\{0,...,HW-1\}$. 
Here we analyze the four different strategies employed for generating the permutations (\autoref{fig:samplingComparison}, top): Random sampling, regular sampling, Halton sampling~\citeapp{halton1964sampling-app}, and plastic sampling~\citeapp{roberts2020plastic-app}.

\begin{figure}%
    \centering%
    \includegraphics[width=\linewidth]{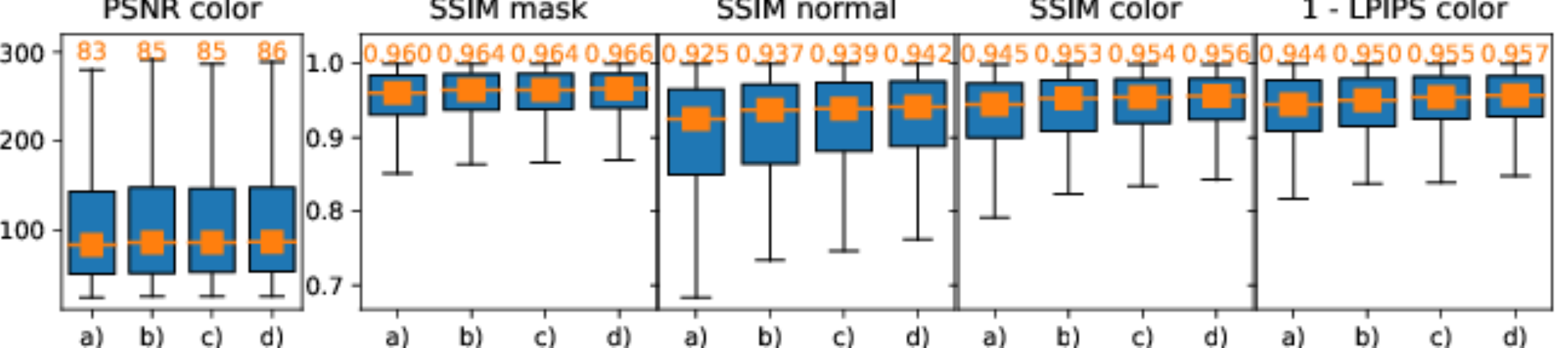}%
    \mycaptionspacing%
    \caption{Reconstruction quality using (a) uniform random, (b) regular, (c) Halton (c), and (c) plastic sampling with $\mu=5\%$ of samples.}%
    \label{fig:results:pattern}%
\end{figure}

Random sampling generates a random permutation of the numbers in $P$. Regular sampling arranges the pixels in a quad-tree and enumerates them using breath-first traversal to generate the sampling pattern. Random and regular sampling introduce, respectively, largely varying sample densities and a strong bias of the sample distribution towards the top of the image.
Both Halton and plastic sampling are deterministic and produce quasi-random sequences with a fairly uniform distribution. As revealed by the quantitative analysis in \autoref{fig:results:pattern}, even though all sampling strategies allow reconstructing the final image at high accuracy, slight differences are noticeable.
Halton and plastic sampling lead to superior quality, in particular w.r.t. the variance of the quality metrics. Plastic sampling, designed as a low-discrepancy sampling sequence, shows the lowest variance and slightly higher scores than Halton sampling. We therefore use plastic sampling in our implementation.

\section{Application to DVR}\label{app:dvr}
In this section, we provide additional details on how the proposed adaptive sampling pipeline is applied to DVR images, as mentioned in \autoref{sec:dvr}.
First we present the changes to the pipeline in terms of input and output channels and the used loss function. Second, we describe how to generate the training data including sampling of transfer functions.

\paragraph*{Input Channels and Loss Function}
First, the input channels to the network pipeline are reinterpreted. For isosurfaces, a mask, normals and depth were passed to the network as input (5 channels per pixel), now color images from the DVR images, together with alpha, depth and normal maps are used as input (8 channels per pixel). The network also only reconstructs color images in RGB$\alpha$ space.

For DVR, depth and normal maps are computed by treating the screen space depth and normal at each sample in object space like a regular color and blended as such with the opacity given by the transfer function (TF). The result is a single depth and normal value per ray which can be interpreted as a weighted average of the depth and normal of all samples along the ray.
We found that adding depth and normals as input channels improves the quality of the reconstruction as it provides additional locally consistent information about the curvature of the object. 

Second, the loss functions on the individual channels as used for isosurfaces are replaced by losses only on the RGB$\alpha$-color.
We apply $L_1$ losses on the color and alpha and an additional LPIPS metric \citeapp{zhang2018perceptual-app} as a perceptual loss, weighted equally:
\begin{equation}
    \mathcal{L}_{\text{dvr}} = \mathcal{L}_{1,rgba} + \mathcal{L}_{\text{LPIPS}, rgb} .
\end{equation}
We found that adding a perceptual loss is critical in reconstructing fine details and sharp silhouettes.
The networks operates in RGB space, other colorspaces like HSV, XYZ or CIELAB did not improve the result. Furthermore, the training data is augmented by randomly shuffling the RGB channels. This helps the network to not overfit for a specific color.

\paragraph*{Data Set Generation}
For training and validation, random transfer functions (TFs) are generated (see below) and Ejecta was used as data set. The test images in the result section use user-generated TFs.
Note that since the low-resolution input for the importance network is also generated with a TF, the network can learn to select features specific to that TF, even though it was never seen during training.

To generate meaningful TFs, first a density histogram is computed and then a Gaussian Mixture Model (GMM) is used to cluster densities in an unsupervised manner. GMMs have been previously used to cluster two-dimensional feature points \citeapp{wang2011}, e.g., density and gradient magnitude. Our approach follows the same idea to cluster one-dimensional feature points, i.e., density values. The GMM represents each cluster as a 1D Gaussian function with a certain mean, i.e., the cluster center, and standard deviation, i.e., the cluster spread. To determine the number of components of the GMM, several GMMs with different numbers of components are build and the one with the lowest Bayesian information criterion (BIC) \citeapp{Schwarz1978EstimatingTD} value is selected. BIC penalizes the number of components and prevents overfitting using many components. 

\begin{figure}[h]
    \centering
    \includegraphics[width=\linewidth]{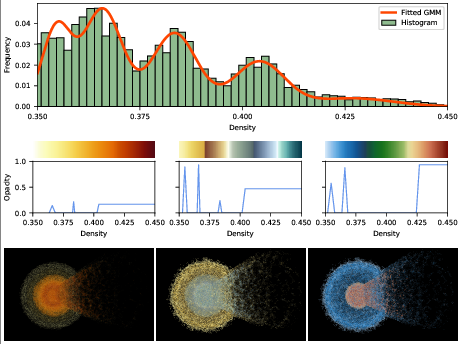}%
    \mycaptionspacing%
    \caption{First row: Histogram of the density values of the Ejecta data set and matched GMM. Second row: three sampled transfer functions with opacity and color. Third row: Renderings from the training data set with those transfer functions.}
    \label{fig:dvr:tfs}
\end{figure}

After computing the GMM, the number of peaks of the TF is sampled uniformly between 3 and 5. The represented density for each peak is sampled from the computed GMM. Next, a width in density space is sampled uniformly from $[0.005, 0.03]$ and the opacity at that peak from $[0.1, 1.0]$.
As colormaps, predefined colormaps from SciVisColor\footnote{\href{https://sciviscolor.org/home/colormaps/}{https://sciviscolor.org/home/colormaps}} are randomly sampled.
The generation process is visualized in \autoref{fig:dvr:tfs}.

We note that it is important for the quality of the reconstruction that the color transfer functions in the training data include a broad range of colors. For example, if the training data only contains desaturated colors, strongly saturated colors during testing cannot be reconstructed.

\section{Pull-Push Algorithm}\label{app:pullpush}
As a baseline method to interpolate the sparse samples, we apply a variation of the push-pull algorithm~\citeapp{gortler1996lumigraph-app,kraus2009pull-app}, see \autoref{alg:Inpainting} for the pseudo code. The algorithm builds upon the idea of mip-map levels: first, the image is downscaled using bilinear interpolation with weights based on the mask. Then, the image is upscaled again and blended with the values at the finer levels with the mask values at the finer levels. We refer to \autoref{sec:method:reconstruction} for more details in the context of the adaptive sampling pipeline.
The pull-push algorithm can be directly extended to fractional masks as shown in \autoref{alg:Inpainting}. During the upsampling stage, the mask is not treated binary, i.e. either take the original pixel at the fine level or use the interpolated value from the coarse level, but fractional with a linear interpolation between the original value and the interpolated value.
Furthermore, the algorithm consists only of linear pooling and interpolation layers which are easy to differentiate with respect to the input mask. We refer to \autoref{app:adjoint:pullpush} for an outline on how to derive the backward pass.

\begin{algorithm}[h]
	\small
\begin{algorithmic}[1]
\Function{Inpainting}{maskIn : HxW, dataIn : HxWxC}
	\If{$\text{H}\leq 1\text{ or }\text{W}\leq 1\text{ or all pixels are filled}$}
		\State \textbf{return} maskIn, dataIn \Comment{end of recursion}
	\EndIf
	\Statex {\em weighted area downsampling:}
	\State maskLow, dataLow = zeros of shape $\frac{H}{2}\text{x}\frac{W}{2}$, $\frac{H}{2}\text{x}\frac{W}{2}\text{x}C$
	\For{$i,j \in \{0,...,\frac{H}{2}-1\}\times\{0,...,\frac{W}{2}-1\}$}
		\State $N_{\text{max}}, N_{\text{avg}}, d = \mathbf{0}^C$
		\For{$a,b \in \{2i,2i+1\}\times\{2j,2j+1\}$} \Comment{loop over neighbors in the fine grid}
			\State $N_{\text{max}} = \max\{N_{\text{max}}, \text{maskIn}[a,b]\}$
			\State $N_{\text{avg}} \pluseq \text{maskIn}[a,b]$
			\State $d \pluseq \text{maskIn}[a,b] \cdot \text{dataIn}[a,b,:]$
		\EndFor
		\If{$N_{\text{avg}}>0$}
			\State $\text{maskLow}[i,j]=N_{\text{max}}$
			\State $\text{dataLow}[i,j,:]=d/N_{\text{avg}}$
		\EndIf
	\EndFor
	
	\Statex {\em recursion:}
	\State maskLow, dataLow = \Call{Inpainting}{maskLow, dataLow}
	
	\Statex {\em weighted bilinear upsampling:}
	\State maskOut, dataOut = zeros of shape $H\text{x}W$, $H\text{x}W\text{x}C$
	\For{$a,b \in \{0,...,H-1\}\times\{0,...,W-1\}$}
		\State $N, W=0, d=\mathbf{0}^C$
		\State $\hat{a}=a\div 2, \hat{b}=b\div 2$ \Comment{Integer-division (round down), indices on the coarse grid}
		\State $a',b' = -1\text{ if }a,b\text{ is even else }+1$
		\State $N=\left\{(\hat{a},\hat{b},\frac{9}{16}), (\hat{a}+a',\hat{b},\frac{3}{16}), (\hat{a},\hat{b}+b',\frac{3}{16}), (\hat{a}+a',\hat{b}+b',\frac{1}{16}) \right\}$
		\For{$(i,j,w)\in N\cap\text{image}$}\Comment{loop over neighbors if within bounds}
			\State $N \pluseq w \text{ maskLow}[i,j]$
			\State $d \pluseq w \text{ maskLow}[i,j] \cdot \text{dataLow}[i,j,:]$
			\State $W \pluseq w$
		\EndFor
		\State $\text{maskOut}[a,b]=\text{maskIn}[i,j], \text{dataOut}[a,b,:]=\text{maskIn}[i,j] \cdot \text{dataIn}[i,j,:]$
		\If{$N>0$}\Comment{blend interpolated values with original data}
			\State $\text{maskOut}[a,b] \pluseq (1-\text{maskIn}[i,j]) \ N / W$
			\State $\text{dataOut}[a,b,:] \pluseq (1-\text{maskIn}[i,j]) \ d / N$
		\EndIf
	\EndFor
	
	\State \textbf{return} maskOut, dataOut
\EndFunction
\end{algorithmic}

	\caption{Pseudocode of the pull-push algorithm for power-of-two input images (a version handling non-power-of-two inputs and the adjoint code for computing the derivative with respect to the mask and data are provided in the source code).}
	\label{alg:Inpainting}
\end{algorithm}

\section{Differentiation of the Sampling and Reconstruction Stages}\label{app:adjoint}

The adjoint code for the gradient propagation in the backward pass is automatically generated by PyTorch for the networks, the loss functions and for the sampling function~\autoref{eq:samplingSigmoid}. For the pull-push algorithm (\autoref{alg:Inpainting}), the adjoint code was manually derived and implemented as a custom operation.
In this section we provide the fundamentals of the adjoint method to manually derive the adjoint code and show how it can be applied to the sampling function and the pull-push algorithm.

\subsection{Fundamentals of the Adjoint Method}
The adjoint method has a long history in Optimal Control Theory, we refer the interested reader to the book by Lions~\citeapp{lions1971optimal} for a complete mathematical introduction. Here, we briefly sketch the fundamentals following the notation by McNamara~\etAl~\citeapp{mcnamara2004fluid}.

By ignoring applications to linear systems and differential equations and focussing on chained functions instead, the adjoint method simplifies to an application of the chain rule.
Let the algorithm be defined as a concatenation of functions $f_i$ with parameters $w_i$ starting from an input value $x_0$,
\begin{equation}
    \begin{aligned}
        x_1 &= f_1(x_0, w_1) \\
        x_2 &= f_2(x_1, w_2) \\
         & \vdots \\
        x_n &= f_n(x_{n-1}, w_n) \\
        s &= J(x_n, w_J) .
    \end{aligned}
    \label{eq:adjointForward}
\end{equation}
The result $s$ has to be a scalar value, this is crucial for the application of the adjoint method in this simple form. 
In the context of neural networks, $x_0$ would be the input image, $f_1$ to $f_n$ the network layers with weights $w_i$ and feature vectors $x_i$, $J$ would be the loss function with target image $w_J$ and $s$ the scalar score.

During training, we are interested in the derivatives $\frac{\partial J}{\partial w_i}$ to update the weights or in e.g. $\frac{\partial J}{\partial x_0}$ to update the initial image in a feature-visualization context.
First, given the -- possibly vector valued -- variables $x_i$ and $w_i$, let the \emph{adjoint variables} $\hat{x}_i$ and $\hat{w}_i$ be defined as the gradient of $s$ with respect to $x_i$ and $w_i$, $\hat{x}_i:=\nabla_{x_i}s, \hat{w}_i:=\nabla_{w_i}s$ as column vectors.
Next, we drop the index $i$, as we require it to index the elements in the input and output vectors, and look at a single function $f\in\R^{N}\times\R^{W}\rightarrow\R^{M}$ with inputs $x\in\R^{N}, w\in\R^{W}$ and output $y\in\R^{M}$. The adjoint variables are then computed using
\begin{equation}
    \hat{x} = J_{f,x}^T(x,w) \hat{y} \ , \ \ 
    \hat{w} = J_{f,w}^T(x,w) \hat{y} .
    \label{eq:adjoint:jacobian1}
\end{equation}
Here, the Jacobian matrix with respect to the different inputs is used, defined as
\begin{equation}
    \left(J_{f,x}\right)_{ij}:=\frac{\partial f_i}{\partial x_j} \ , \ \ 
    \left(J_{f,w}\right)_{ij}:=\frac{\partial f_i}{\partial w_j} .
    \label{eq:adjoint:jacobian2}
\end{equation}

As one can see, given the adjoint variable of the output $\hat{y}$, the adjoint method propagates these gradients back through the derivatives of $f$ to the adjoint variables of the inputs $\hat{x}$ and $\hat{w}$.
In the context of the chained function~\autoref{eq:adjointForward}, this implies that, starting with gradients on the output $\hat{x}_n$ from the loss function, gradients are first propagated to $\hat{x}_{n-1}, \hat{w}_n$ via $J_{f_i}$, then to $\hat{x}_{n-2}, \hat{w}_{n-1}$, and so on until $\hat{x}_0, \hat{w}_1$ is reached.

To provide custom differentiable operations, two functions have to be provided:
first, the forward code $y\leftarrow f(x,w)$ with input $x$ and parameter $w$, and
second, the backward code to compute $\hat{x}$ and $\hat{w}$ from $\hat{y}$, possibly using $x,w$ from the forward pass again to compute the Jacobian.

\subsection{Backward Pass of the Sampling Function}
Using the theory above, we now present the adjoint code for the differentiable sampling from \autoref{sec:method:sampling}. This serves to highlight what is differentiated and how the gradients are propagated. Note that these functions are implemented based on PyTorch functions, PyTorch can automatically compute the derivatives.

The differentiable sampling stage takes the importance map $I$ as input and produces the image of sparse samples $S$. In the framework of \autoref{eq:adjointForward}, this can be seen as block of functions that is cut out in the middle. As parameters, the target mean $\mu$ and lower bound $l$, the sampling steepness $\alpha$, the sample pattern $P$ and the target image $T$ are used. Note that no optimization with respect to these parameters is performed, their respective adjoint variables are unused.
To recapitulate, the sampling is performed using the following steps:
\begin{subequations}
  \begin{align}
     \mu_I &= \frac{1}{WH}\sum_{ij}{I_{ij}} \ , \ \ I^{(1)}=I \\
     I'_{ij} &= \operatorname{min}\left\{ 1, l+I^{(1)}_{ij}\frac{\mu-l}{\mu_I+\epsilon} \right\} \\
     S_{ij} &= \operatorname{sig}(\alpha(I'_{ij}-P_{ij}))T_{ij} \ \ \text{with} \ \operatorname{sig}(x)=\frac{1}{1+e^{-x}}.
  \end{align}
\end{subequations}
Note that the second and third function act on each pixel $ij$ of the images independently. Therefore, we use them as per-element functions to simplify the notation of the derivatives. Using the matrix notation from \autoref{eq:adjoint:jacobian1}, this would imply a diagonal Jacobian. Furthermore, $T_{ij}$ and $S_{ij}$ return the vector of channels at the specified location.
In order to stay within the presented framework of the adjoint method, if variables are used by a function and later again by another function, these variables are passed through as additional outputs ($I^{(1)}=I$).

For the backward pass, we are given the gradients of the output $\hat{S}$ from the backward pass of the reconstruction. This equates to $\hat{x}_n$ in \autoref{eq:adjointForward}.
Then the gradients are propagated through the sampling algorithm in reverse order:
\begin{subequations}
    \begin{align}
        \phantom{\hat{I'}_{ij}}
        &\begin{aligned}
            \mathllap{\hat{T}_{ij}} &= \operatorname{sig}(\alpha(I'_{ij}-P_{ij})) \hat{S}_{ij} \\
            \mathllap{\hat{P}_{ij}} &= -(\alpha T_{ij} \operatorname{sig}'\left(\alpha (I'_{ij}-P_{ij})\right)^T \hat{S}_{ij} \\
            \mathllap{\hat{I'}_{ij}} &= (\alpha T_{ij} \operatorname{sig}'\left(\alpha (I'_{ij}-P_{ij})\right)^T \hat{S}_{ij} \\
             &\text{with} \ \operatorname{sig}'(x) = \frac{d}{dx}\operatorname{sig}(x)=\operatorname{sig}(x)\operatorname{sig}(-x)
        \end{aligned}
    \\
        \phantom{\hat{I'}_{ij}}
        &\begin{aligned}
            \mathllap{\hat{I}^{(1)}_{ij}} &= \begin{cases}
                \frac{\mu-l}{\mu_I+\epsilon} \hat{I}'_{ij} \ , \ I'_{ij} < 1 \\
                0 \ , \ I'_{ij}\geq 1
                \end{cases} \\
            \mathllap{\hat{\mu}_l} &= \sum_{ij}{\left(\begin{cases}
                \frac{I^{(1)}_{ij}(l-\mu)}{(\mu_l+\epsilon)^2} \hat{I}'_{ij} \ , \ I'_{ij} < 1 \\
                0 \ , \ I'_{ij}\geq 1
                \end{cases}\right)} \\
            & \text{derivatives for }\mu, l, \epsilon\text{ are omitted}
        \end{aligned}
    \\
        \phantom{\hat{I'}_{ij}}
        &\begin{aligned}
            \mathllap{\hat{I}_{ij}} &= \hat{I}^{(1)}_{ij} + \frac{1}{WH}\hat{\mu}_l
        \end{aligned}
    \end{align}
\end{subequations}

\subsection{Backward Pass of the Pull-Push Algorithm}\label{app:adjoint:pullpush}

As one can see in the previous section, deriving the adjoint code is done mechanically by deriving each line of code with respect to the inputs. This, however, produces a vastly longer code, therefore, we only outline the steps to derive the adjoint code of the pull-push algorithm \autoref{alg:Inpainting}. The full source code is available in the online repository.

The algorithm is a recursive algorithm with three stages: the downsampling to the coarse level, the recursive call, and the upsampling and interpolation with the fine level.
During the backward pass, the order is reversed. First, the adjoint of the upsampling and interpolation at the finest level. Then the adjoint of the recursive call, which itself is the adjoint of upsampling, recursion and downsampling. And lastly the adjoint of the downsampling.

\bibliographystyleapp{abbrv-doi-narrow}
\bibliographyapp{main}

\end{document}